\newcommand{\pT}{$p_\mathrm{T}$}
\newcommand{\pp}{$pp$}
\newcommand{\PbPb}{Pb+Pb}
\newcommand{\pPb}{$p$+Pb}
\newcommand{\RAA}{$R_\mathrm{AA}$}
\newcommand{\RAAP}{$R_\mathrm{AA}^\mathrm{pair}$}
\newcommand{\SNN}{$\sqrt{s_\mathrm{NN}}$}
\newcommand{\gajet}{$\gamma$-jet}
\newcommand{\xJ}{$x_\mathrm{J}$}
\newcommand{\xJg}{$x_\mathrm{J\gamma}$}
\newcommand{\pl}{path length}
\newcommand{\IAA}{$I_\mathrm{AA}$}
\newcommand{\RCP}{$R_\mathrm{CP}$}

\newcommand{\pythia}{\mbox{\textsc{Pythia}8}}

\documentclass{webofc}
\usepackage[varg]{txfonts}   % Web of Conferences font
\usepackage{hyperref}
\usepackage{url}
\usepackage{graphicx}
\usepackage{subcaption}
\usepackage{caption}

\hypersetup{colorlinks=true,citecolor=blue,urlcolor=blue,linkcolor=blue}
\begin{document}
\captionsetup{font=footnotesize,labelfont=footnotesize}
\title{Hard-jet correlations in large and small systems}

\author{\firstname{Riccardo} \lastname{Longo}\inst{1}\fnsep\thanks{\email{riccardo.longo@cern.ch}} 
        % etc.
}

\institute{University of Illinois at Urbana-Champaign, Dept. of Physics, 1110 W Green St., 61801, Urbana (IL)
          }
\abstract{Hard-jet correlations probe parton energy loss and the microscopic structure of the quark-gluon plasma formed in ultra-relativistic heavy-ion collisions. The correlation of high-$p_\mathrm{T}$ jets with other jets, hadrons, or electroweak bosons, offers differential sensitivity to medium-induced effects such as momentum broadening, color decoherence, and medium response in different types of nuclear reactions. Such correlations can also be used to study cold nuclear matter effects arising in $p$+A collisions. This proceeding summarizes recent advances achieved by studying hard-jet correlations in large and small systems discussed at Hard Probes 2024, complementing the experimental jet overview~\cite{Yaxian}.}
\maketitle
\section{Introduction}
\label{intro}
Ultra-relativistic heavy-ion (HI) collisions at the CERN LHC and RHIC recreate droplets of Quark-Gluon Plasma (QGP) in the laboratory. The QGP exhibits strong collective phenomena and significant modifications of hard probes, providing a unique environment to study the emergent properties of QCD in the non-perturbative regime.

A central goal of HI physics is to connect the macroscopic collective behavior of the QGP with its microscopic structure, governed by quarks and gluons. Jets are ideal probes of this structure, produced early in the collision and traversing the medium throughout its evolution. Their energy loss depends on the path length and initiating parton flavor, while the QGP response can modify soft particle production along and opposite to the jet direction.

A full understanding of the QGP’s evolution also requires a precise characterization of the initial state, essential to disentangle cold nuclear effects from medium-induced phenomena.

In this talk, I reviewed recent hard-jet correlation measurements, including jets, hadrons, and bosons, in large (A+A) and small ($p/d$+A, \pp) systems. Focusing on key questions, I highlighted new insights from recent results.

\section{Large systems} 
\vspace{-0.8cm}
\subsection{How does the quenching depend on jet properties?} 
In HI collisions, jet quenching is expected to depend on the initiating parton's flavor. In particular, gluon-initiated jets (with color charge $C_\mathrm{g}=3$) are expected to lose more energy in the medium than quark-initiated jets (color charge $C_\mathrm{q}=4/3$) due to their larger color charge and higher medium-induced radiation probability. Perturbative QCD energy-loss calculations thus predict roughly a factor of $C_\mathrm{g}/C_\mathrm{q}\sim 2.25$ more radiative energy loss for a gluon jet than a quark jet under identical conditions \cite{Caron-Huot:2008zna}. This prediction can be tested by comparing the nuclear modification factor (\RAA) of boson-tagged jets, which are predominantly quark-initiated, to that of inclusive jets, which have a larger fraction of gluon-initiated jets. 

\begin{figure}[h]
\centering
\begin{subfigure}{0.38\textwidth}
    \includegraphics[width=\linewidth,clip]{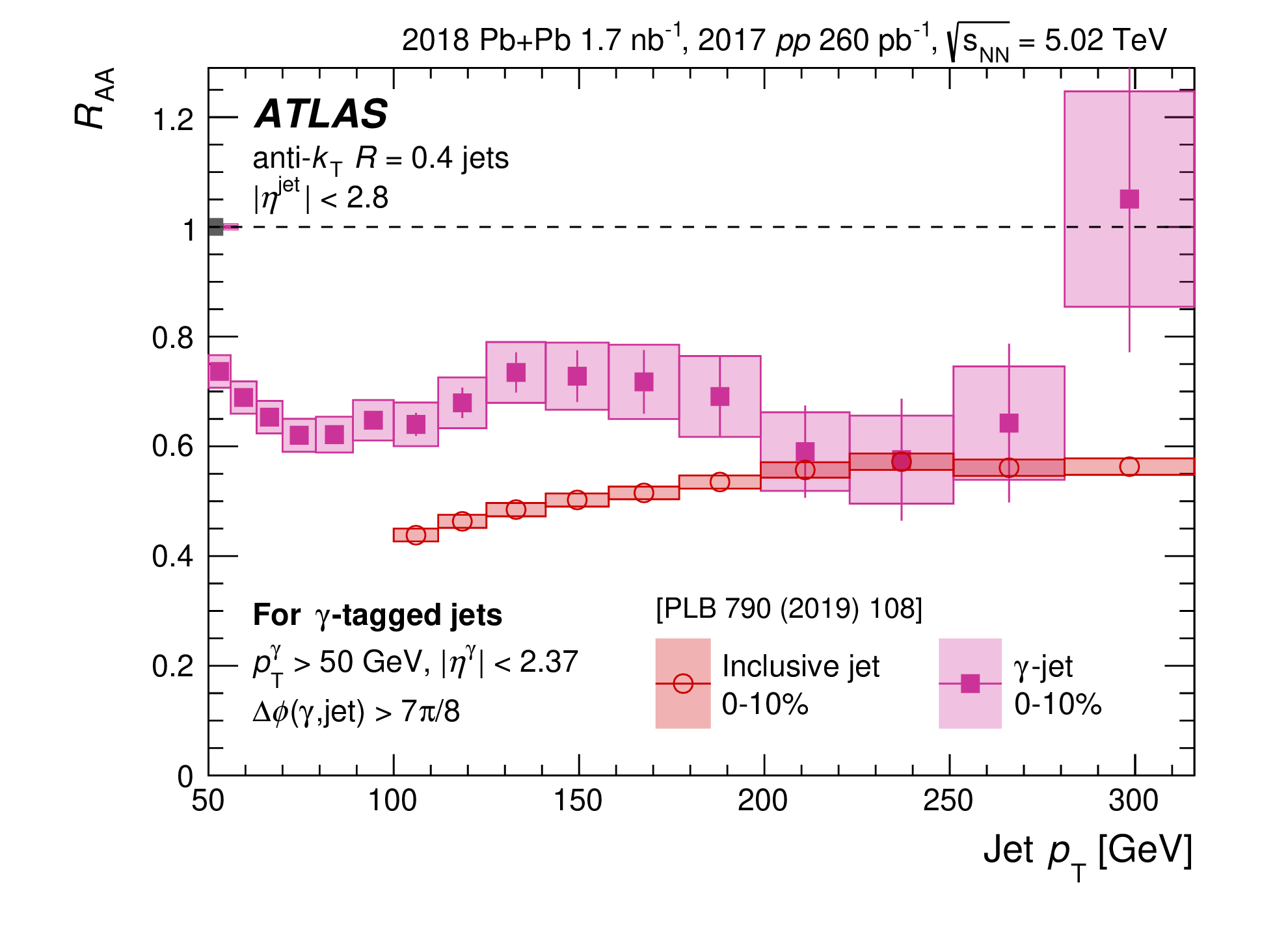}
    \caption{}
    \label{fig:fig1b}
\end{subfigure}
\hfill
\begin{subfigure}{0.23\textwidth}
    \includegraphics[width=\linewidth,clip]{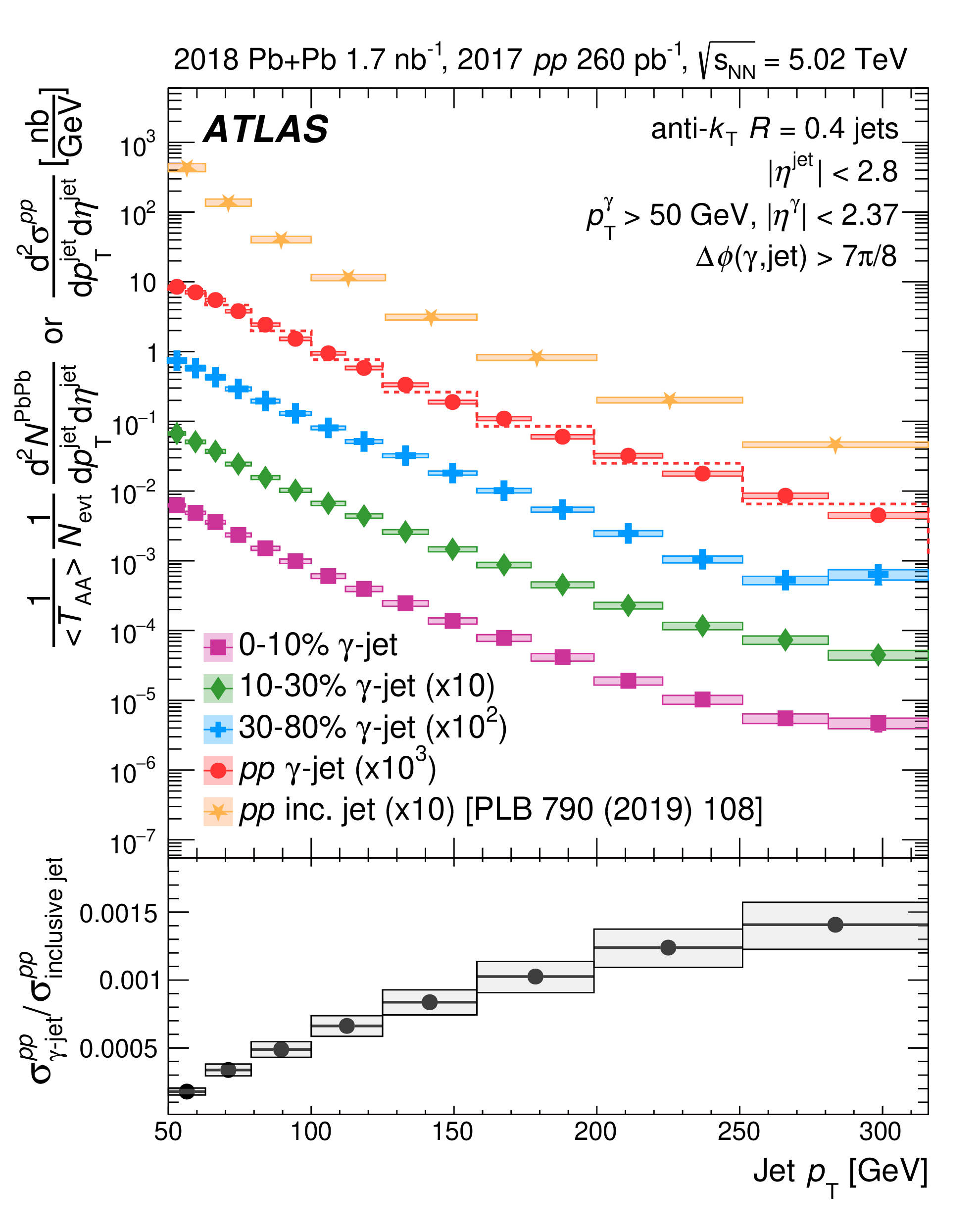}
    \caption{}
    \label{fig:fig1a}
\end{subfigure}
\hfill
\begin{subfigure}{0.31\textwidth}
    \includegraphics[width=\linewidth,clip]{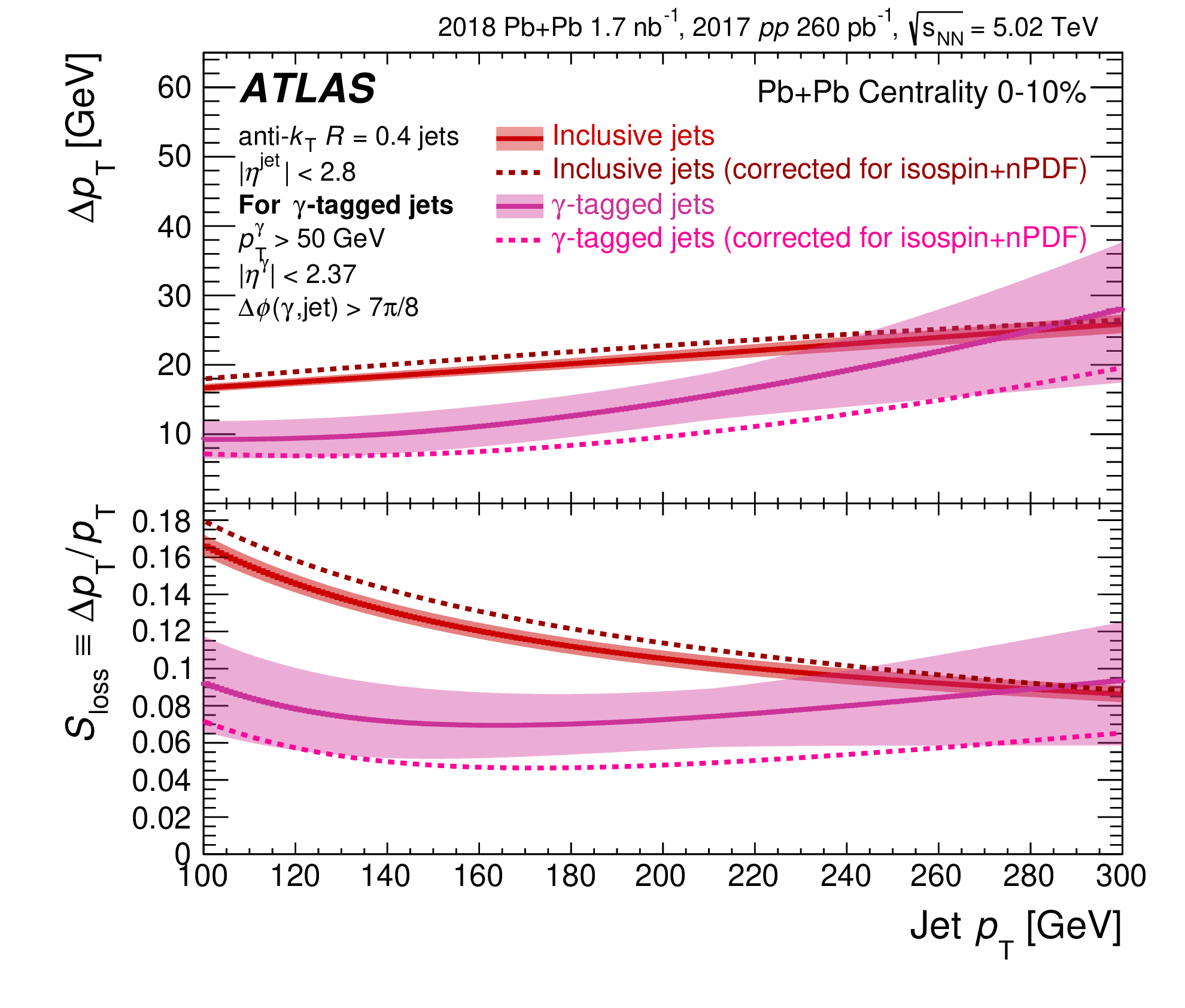}
    \caption{}
    \label{fig:fig1c}
\end{subfigure}
\vspace{-0.5cm}
\caption{\footnotesize (a) The \RAA\, of photon-tagged jets (filled squares) as a function of \pT\, for 0-10\% Pb+Pb events, overlaid with that of inclusive jets (open circles) in the same centrality range \cite{ATLAS:2023iad}. (b)  Yields of photon-tagged jets as a function of \pT\, in Pb+Pb  events for different centrality classes and the differential cross-section in pp events (circles). (c) ATLAS $S_\mathrm{loss}$ extraction for inclusive and $\gamma$-tagged jets.  }
\label{fig-1}
\end{figure}

ATLAS recently reported a study of the \RAA\ of inclusive jets and photon-tagged jets in central (0--10\%) Pb+Pb collisions at \SNN=5.02~TeV, using the full Run 2 statistics \cite{ATLAS:2023iad}. From the results of this comparison (see Fig.~\ref{fig:fig1b}), the \RAA\ of $\gamma$-tagged jets is significantly higher than the one for inclusive jets in the \pT\ range below 200~GeV. Above this value, the results become compatible within experimental uncertainties. 

While interpreting these results, one has to remember that the \RAA\ intrinsically depends on the shape of the initial production spectrum, with steeper spectra resulting in a lower \RAA\ for the same magnitude of energy loss. As shown in Fig.~\ref{fig:fig1a}, inclusive jets in \pp\ are characterized by a steeper \pT\ spectrum compared to \gajet s. To characterize the energy loss with reduced sensitivity on the \pT\ spectral shape, ATLAS has analyzed the fractional energy loss, $S_\mathrm{loss}$ \cite{Brewer:2018dfs} for both jet samples. The results, corrected for isospin and nuclear PDF effects, are shown in Fig.~\ref{fig:fig1c}. The \gajet s $S_\mathrm{loss}$ is significantly lower than that for inclusive jets, predominantly gluon-initiated. These data provide the strongest confirmation to date of larger jet quenching for gluon jets compared with quark jets.
\vspace{-0.3cm}

\begin{figure}[h]
\centering
\begin{subfigure}{0.25\textwidth}
    \includegraphics[width=\linewidth,clip]{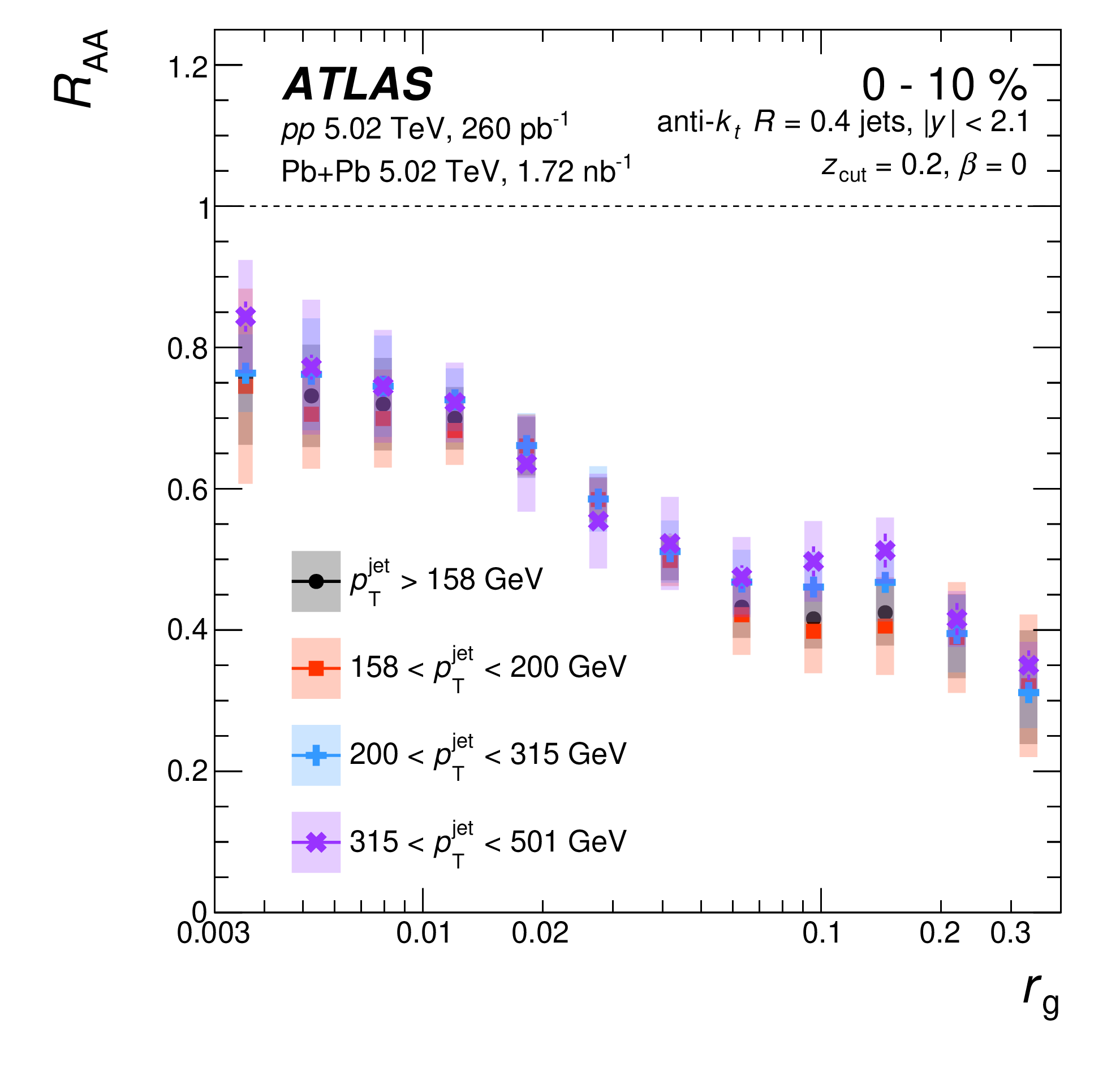}
    \caption{}
    \label{fig:fig2a}
\end{subfigure}
\hfill
\begin{subfigure}{0.33\textwidth}
    \includegraphics[width=\linewidth,clip]{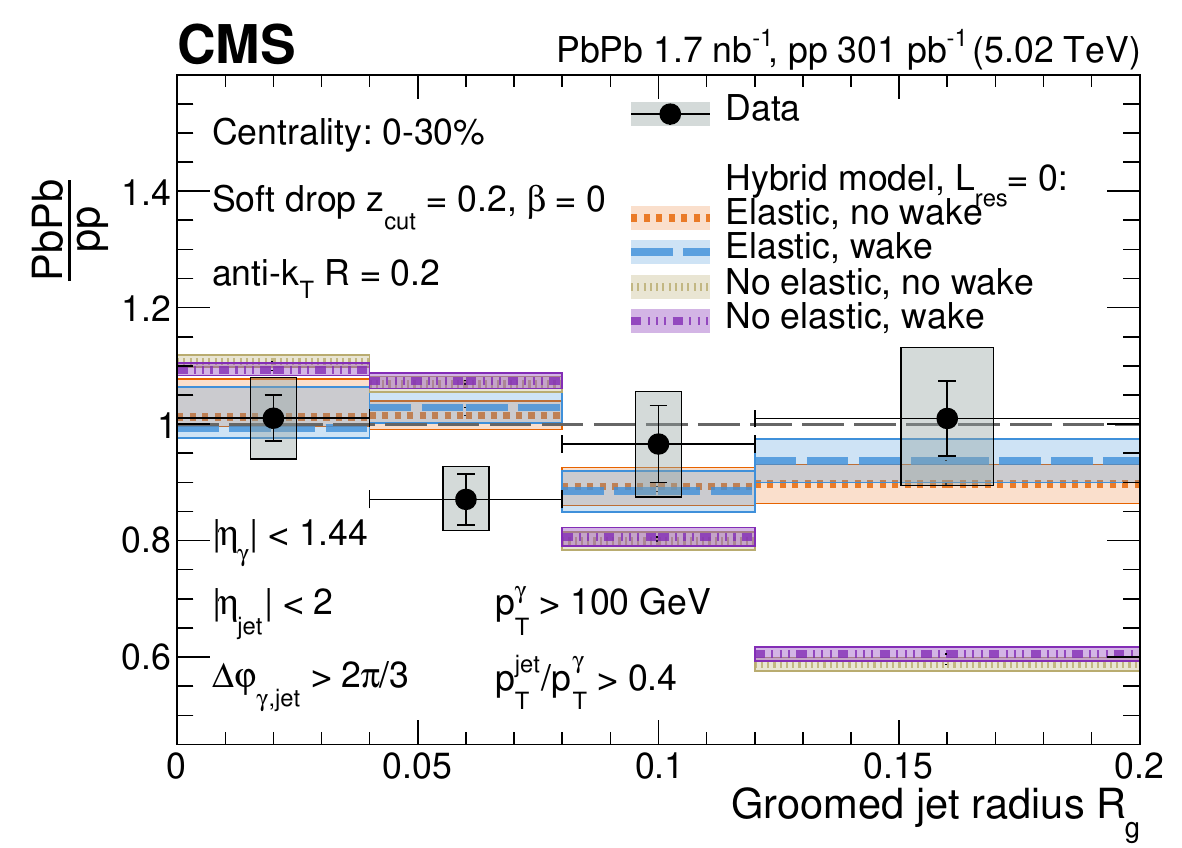}
        \caption{}
    \label{fig:fig2b}
\end{subfigure}
\hfill
\begin{subfigure}{0.33\textwidth}
    \includegraphics[width=\linewidth,clip]{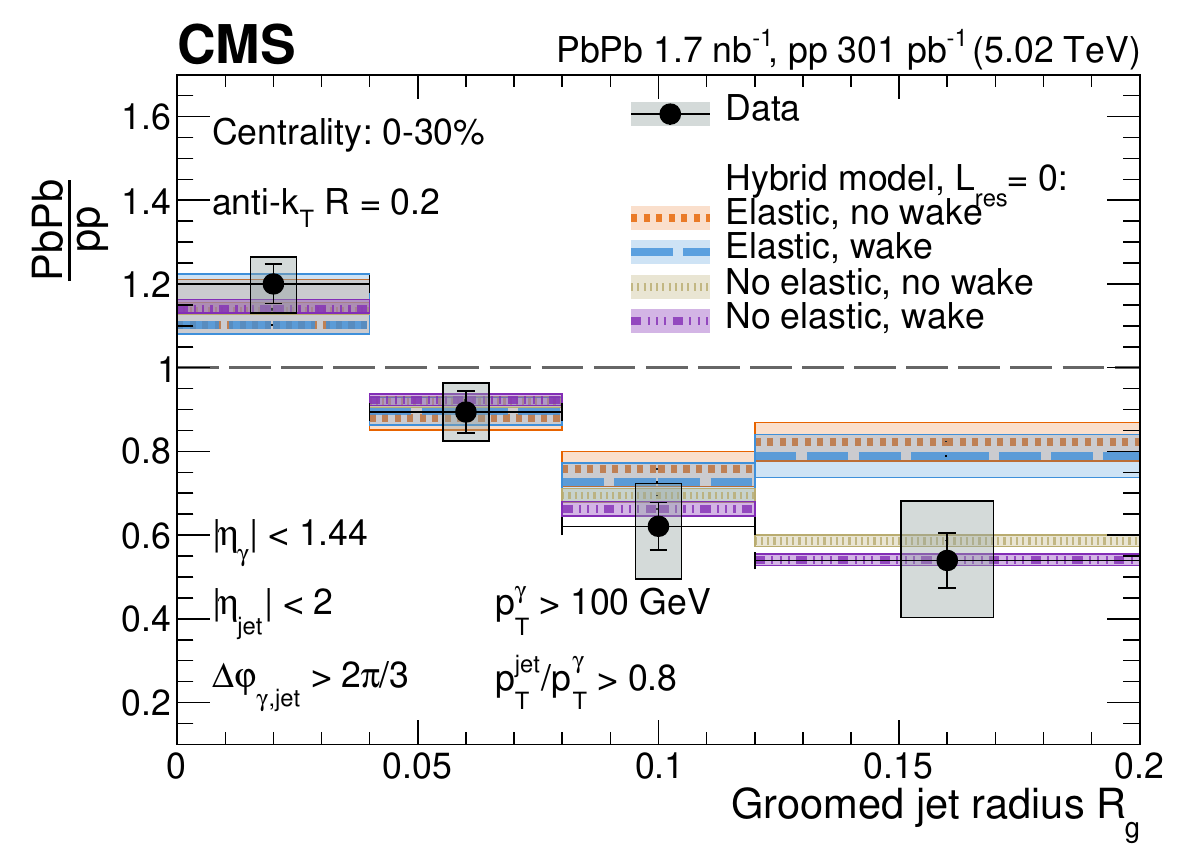}
        \caption{}
    \label{fig:fig2c}
\end{subfigure}
\vspace{-0.5cm}
\caption{\footnotesize(a) The \RAA\, of inclusive jets as a function of the groomed jet radius measured by ATLAS \cite{ATLAS:2022vii} for 0-10\% Pb+Pb events. (b) Ratio of the normalized  \gajet s yields of Pb+Pb to \pp\ data measured by CMS \cite{CMS:2024zjn} as a function of groomed jet radius, for events characterized by large photon-jet imbalance (\xJg\ $>$ 0.4) in 0-30\%Pb+Pb events. (c) Same as (b) but for less quenched jet selections (\xJg\ $>$ 0.8).}
\label{fig-2}
\end{figure}

In recent years, jet substructure has been extensively studied to investigate its modification in the medium and to probe fundamental properties of the quark-gluon plasma, such as the \textit{color coherence} scale~\cite{Casalderrey-Solana:2012evi}. This scale characterizes the distance between color charges below which the QGP cannot resolve small-angle splittings of the parton shower. ATLAS has carried out a series of analyses studying the jet \RAA\ as a function of the configuration of the two hardest prongs within the jet radius\cite{ATLAS:2022vii, ATLAS:2023hso}, exposing a substantial dependence on their angular separation, with narrower jets characterized by lower jet energy loss (see Fig.~\ref{fig:fig2a}). This trend is consistent with a color decoherence picture, where partons within a jet are sufficiently separated to be resolved individually by the medium. 

Nevertheless, the interpretation of \RAA\ results is complicated by the \textit{selection bias} that leads less quenched, narrower jets to be more likely present in the detected sample. The flavor dependence of the energy loss also introduces additional hurdles in directly interpreting the results. The measurement of jet substructure in \gajet s events has been proposed to work around these issues and provide more directly interpretable results. CMS has recently completed the analysis of the girth and groomed radius of $\gamma$-tagged jets \cite{CMS:2024zjn}, presenting the area-normalized ratio of R=0.2 \gajet s yield between Pb+Pb and \pp\  different selections of photon-jet momentum imbalance, \xJg. The results show little suppression for unbalanced configurations, where the jet opposite the photon suffered more quenching in the QGP (Fig.~\ref{fig:fig2b}) compared to less quenched jets (Fig.~\ref{fig:fig2c}), where the narrowing is still observed. Notably, no model can capture the trends displayed by the data with both selections. 
CMS has built on these $\gamma$-tagged results and investigated, similarly, the jet axis decorrelation \cite{MPark} for R = 0.3 jets. 
When comparing CMS \gajet s results to the ATLAS inclusive ones, it is relevant to keep in mind a few relevant differences between the measurements, such as the different flavor implications between inclusive and photon tagged jets, the different jet radii used in the analyses as well as the different observables (\RAA\ vs area normalized ratios). 
\vspace{-0.3cm}

Carrying out an \RAA\ measurement for larger radii in this kind of events, providing a complementary input to the ATLAS inclusive jet substructure analysis, seems the natural next step to move forward the understanding of the interplay between color coherence effects and flavor dependence of energy loss. 

\subsection{How does the energy lost depend on the \pl\ of the jet in the medium?} 
\begin{figure}[h]
\centering
\begin{subfigure}{0.33\textwidth}
    \includegraphics[width=\linewidth,clip]{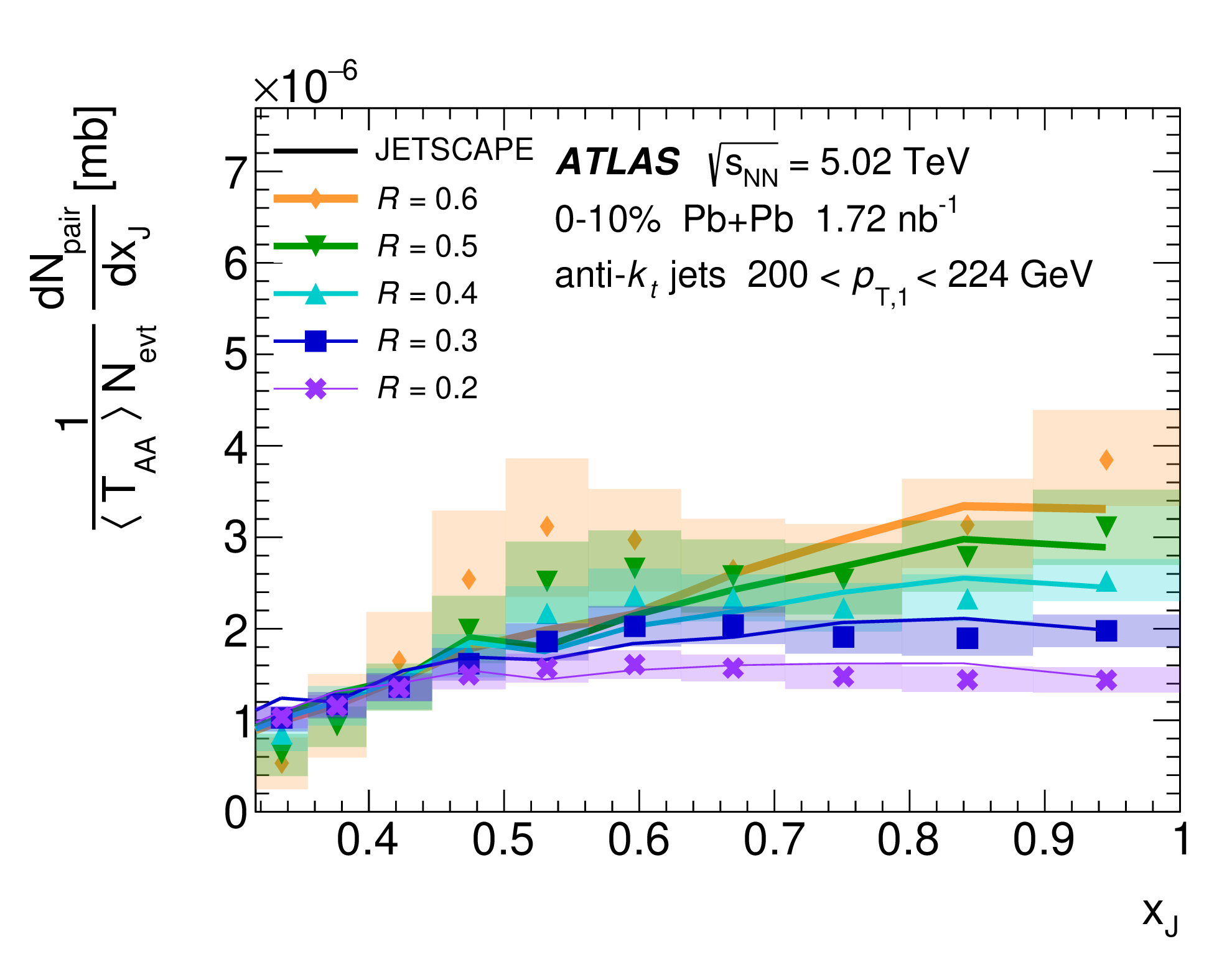}
        \caption{}
    \label{fig:fig3a}
\end{subfigure}
\hfill
\begin{subfigure}{0.33\textwidth}
    \includegraphics[width=\linewidth,clip]{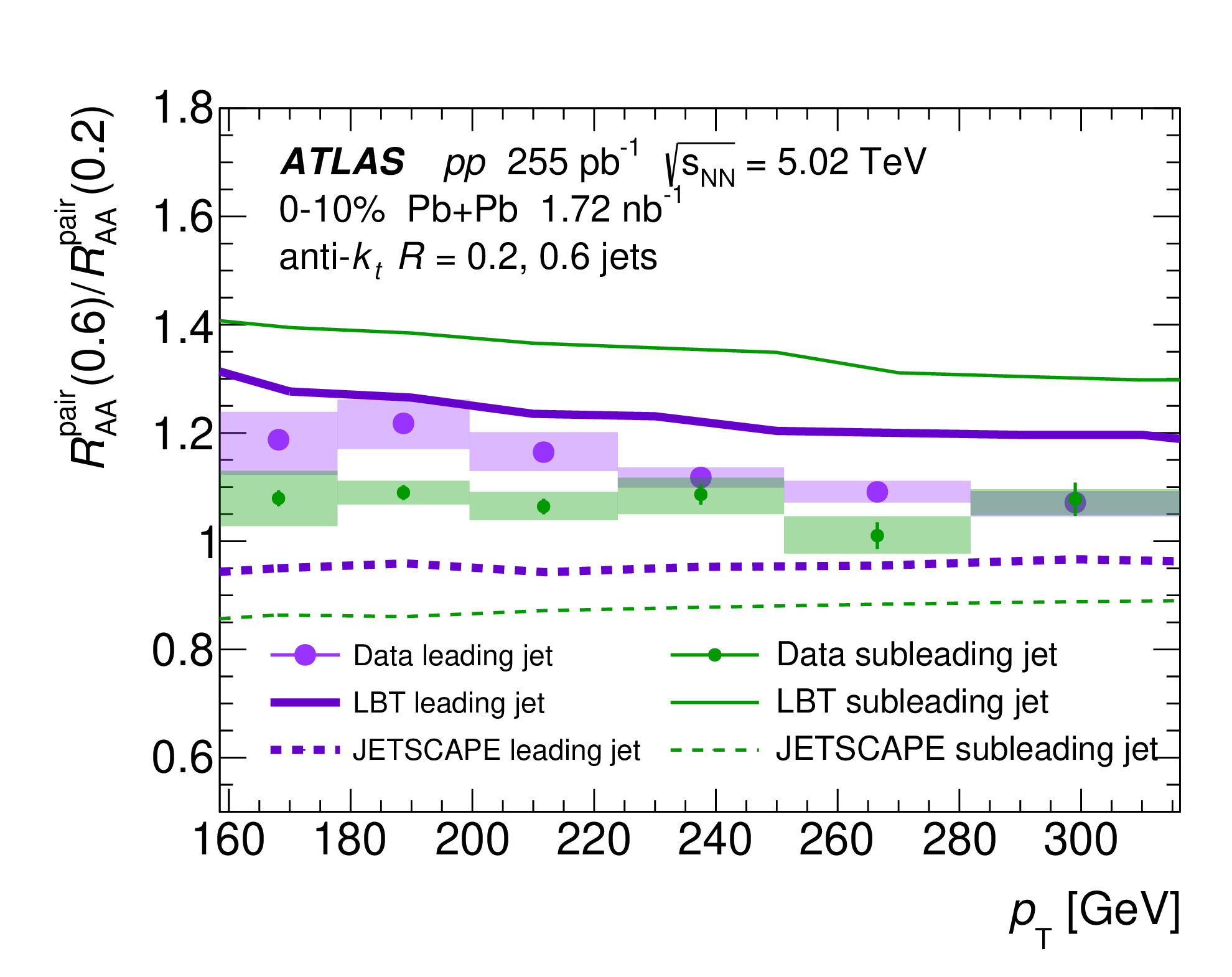}
        \caption{}
    \label{fig:fig3b}
\end{subfigure}
\hfill
\begin{subfigure}{0.27\textwidth}
    \includegraphics[width=\linewidth,clip]{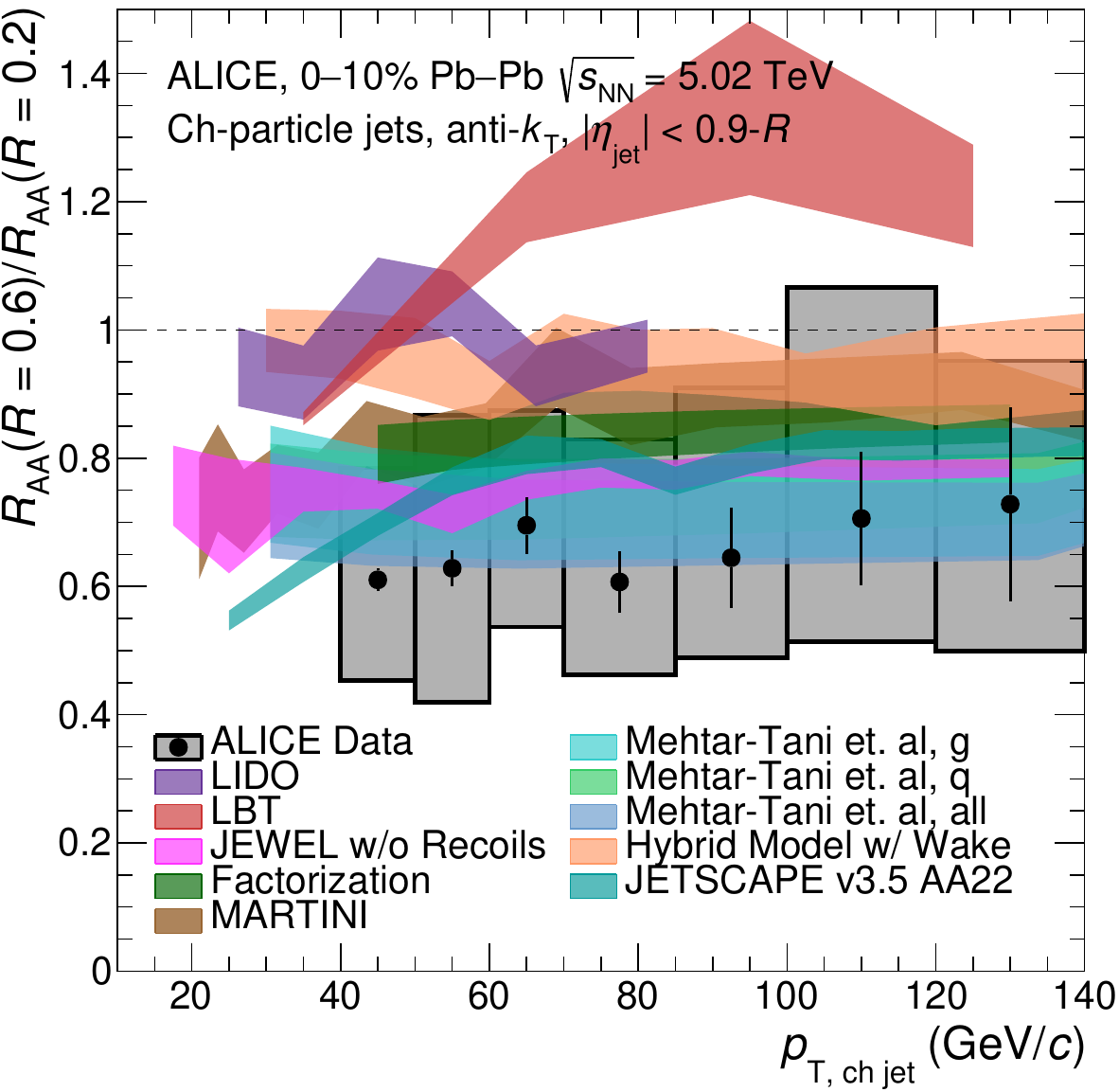}
        \caption{}
    \label{fig:fig3c}
\end{subfigure}
\vspace{-0.5cm}
\caption{\footnotesize (a) ATLAS dijet asymmetry for jets of R=0.2, 0.3, 0.4, 0.5, 0.6, compared with JETSCAPE (LBT+MATTER) predictions (b) Ratio between ATLAS \RAAP\ for large (R=0.6) and small (R=0.2) jet radii, for both leading and subleading jets. The results compared to predictions from LBT and JETSCAPE. (c) Ratio between \RAA\ for large (R=0.6) and small (R=0.2) charged track jets measured by ALICE.}
\end{figure}

The energy lost by jets in the medium is expected to depend not only on flavor and substructure but also on the \pl\ traversed by the jet through the QGP. Dijet pairs offer access to the \pl-dependence of energy loss via their momentum imbalance, \xJ, which reflects the different in-medium path lengths of the two jets. ATLAS has progressively refined the understanding of dijet asymmetry using Run 2 data. A clear asymmetry in \PbPb\ collisions compared to \pp\ was observed for R=0.4 jets~\cite{ATLAS:2022vii}, and the analysis was recently extended to a range of jet radii (R=0.2, 0.3, 0.4, 0.5, 0.6)~\cite{ATLAS:2024jtu}; see Fig.~\ref{fig:fig3a} for the most central \PbPb\ collisions. The results are compared with JETSCAPE (LBT+MATTER) predictions, which fail to capture the observed R-dependence in the mid-\xJ\ region.

ATLAS also measured the dijet nuclear modification factor, \RAAP, by projecting the distributions along the leading and subleading jet \pT, and comparing the results for different radii (Fig.~\ref{fig:fig3b}). Interestingly, larger-R jets appear less suppressed than smaller ones for both leading and subleading jets. Neither LBT nor JETSCAPE accurately describes both the leading–subleading ordering and the R-dependence simultaneously. These findings become even more compelling when compared to ALICE’s radial scan of charged-jet \RAA\cite{ALICE:2023waz}, where R=0.6 jets were more suppressed than R=0.2 jets. While differences in the analyses exist—jets vs. dijets, charged vs. calorimeter jets, and differing rapidity coverage—a unified interpretation of these results is a crucial next step toward fully understanding the mechanisms of parton energy loss.

\begin{figure}[h]
\centering
\begin{subfigure}{0.45\textwidth}
    \includegraphics[width=\linewidth,clip]{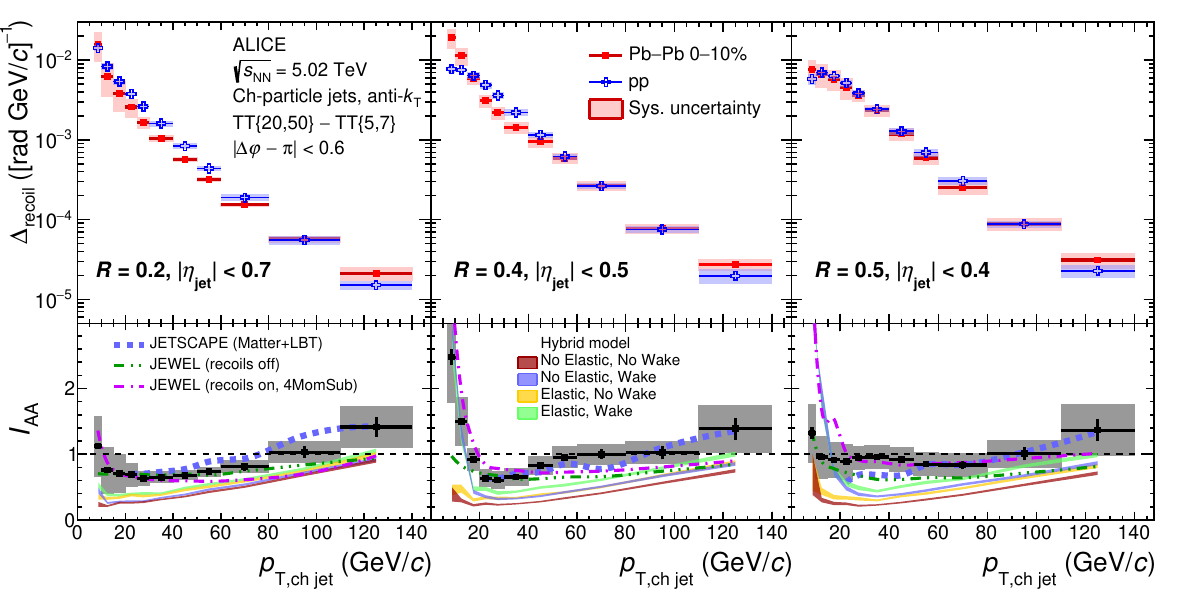}
        \caption{}
    \label{fig:fig4a}
\end{subfigure}
\hfill
\begin{subfigure}{0.46\textwidth}
    \includegraphics[width=\linewidth,clip]{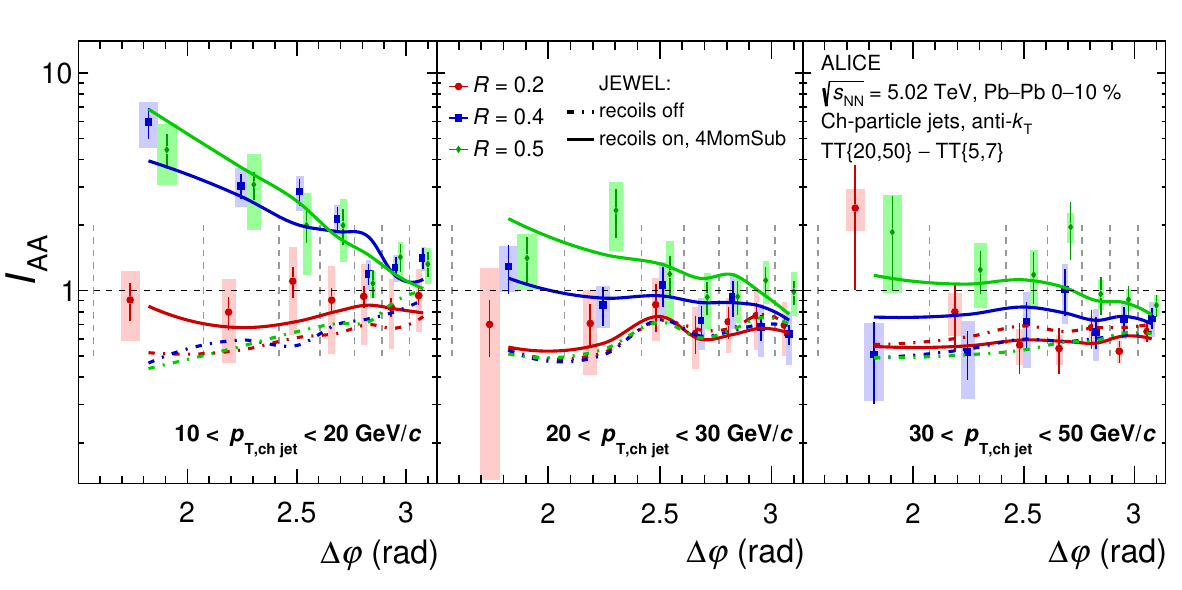}
        \caption{}
    \label{fig:fig4b}
\end{subfigure}
\vspace{-0.5cm}
\caption{\footnotesize(a) Underlying event-corrected \pT\ distributions for \pp\ and central \PbPb\ collisions and corresponding \IAA from ALICE. (b) \IAA\ for jets with R=0.2, 0.4 and 0.5 for different jet \pT\ selections as a function of the azimuthal separation of the trigger hadron and recoil jet, as measured by ALICE.}
\end{figure}

Additional insights into the \pl\ dependence of energy loss can also be obtained through semi-inclusive jet measurements. ALICE recently reported the measurement of the charged-particle jet yield recoiling from high-\pT\ hadrons in both \pp\ and central \PbPb\ collisions~\cite{ALICE:2023qve}. A data-driven statistical method was used to subtract the underlying event contribution in \PbPb\ collisions, enabling the analysis for charged jets with \pT\ greater than 7 GeV. The nuclear modification, studied in this measurement using the \IAA, reveals several intriguing features. First, the \IAA\ increases when going from small to large radii (see Fig.~\ref{fig:fig4a}), an observation qualitatively compatible with the ATLAS dijet measurement~\cite{ATLAS:2024jtu}, suggesting that larger radii recapture soft radiation at larger angles generated by the interaction with the medium. A suppression is observed at low \pT, followed by an enhancement at high \pT, where the \IAA\ exceeds unity. This behavior was analyzed in~\cite{He:2024rcv} and traced back to the quenching experienced by the trigger hadron. Azimuthal broadening was also observed for low-\pT\ jets with R=0.4 and R=0.5 (see Fig.~\ref{fig:fig4b}), an effect that can be interpreted in terms of the wake induced in the medium by the jet. Jets with larger radii tend to incorporate more soft particles from the wake, leading to the observed azimuthal decorrelations.

\vspace{-0.3cm}
\vspace{-0.2cm}

\subsection{What happens to the energy deposited by the jet in the medium?} 
For a comprehensive description of the interaction between jets and the QGP, it is critical to also understand the effects induced in the medium by the jet. When a high-energy parton traverses the QGP, it deposits energy into the medium, inducing an enhancement of soft particle production in the same direction as the jet, commonly referred to as the \textit{wake}. Simultaneously, as the jet moves forward, it depletes strongly interacting matter from its path, creating the so-called \textit{diffusion wake}. The latter is particularly elusive to measure, as it is predicted to correspond, on average, to the loss of less than one particle per unit of pseudorapidity and azimuthal angle~\cite{Yang:2022nei}.

\begin{figure}[h]
\centering
\begin{subfigure}{0.25\textwidth}
    \includegraphics[width=\linewidth,clip]{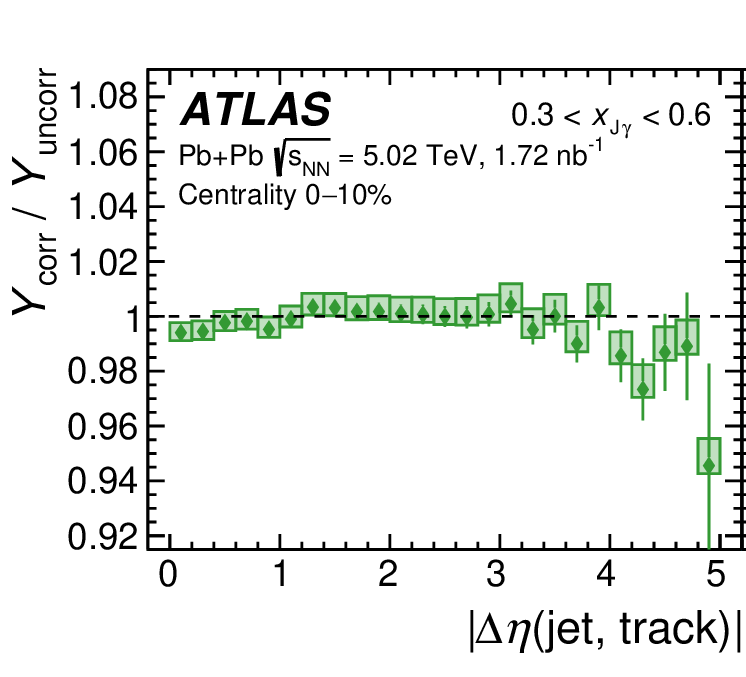}
        \caption{}
    \label{fig:fig5a}
\end{subfigure}
\hfill
\begin{subfigure}{0.42\textwidth}
    \includegraphics[width=\linewidth,clip]{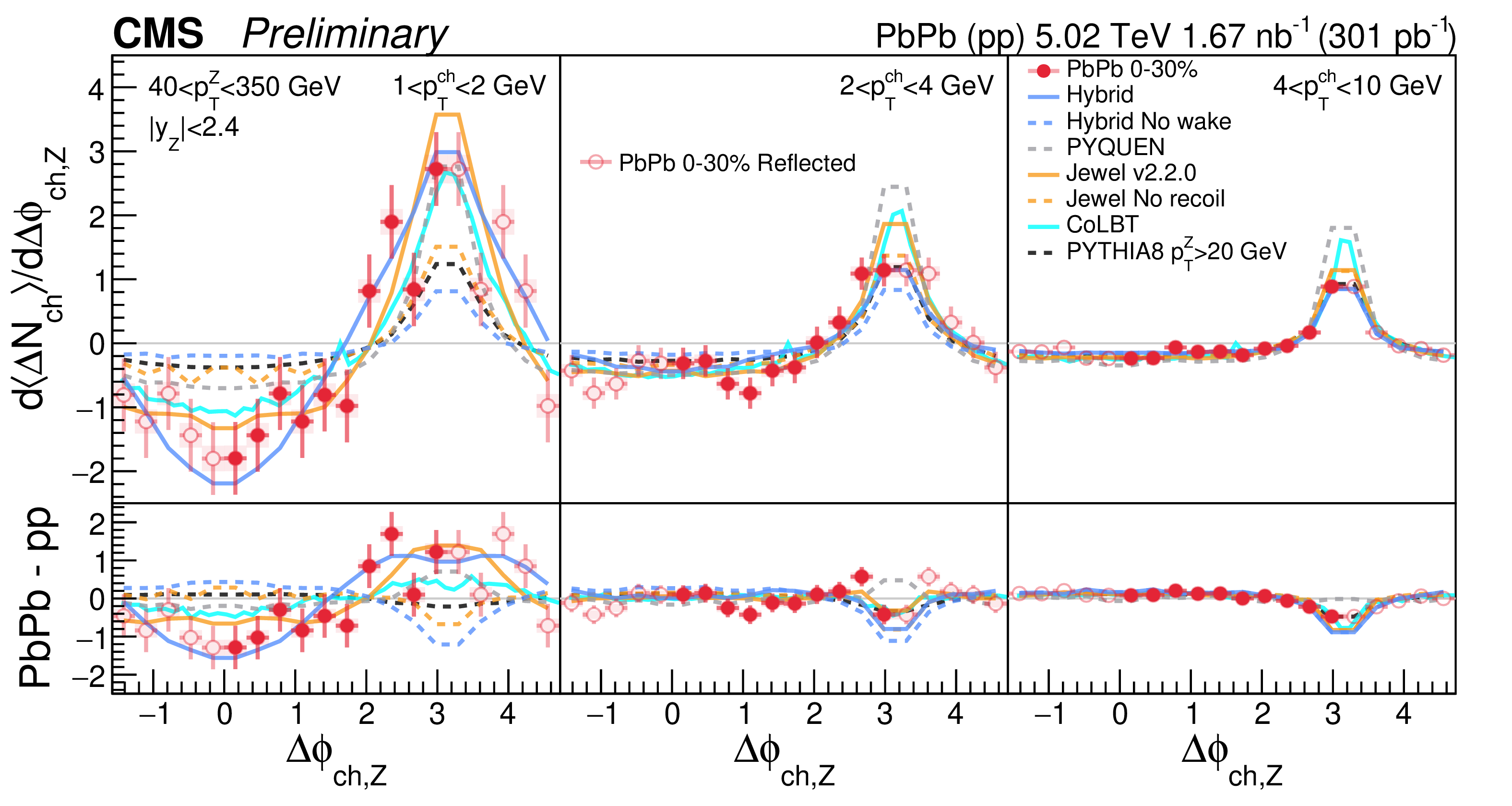}
        \caption{}
    \label{fig:fig5b}
\end{subfigure}
\hfill
\begin{subfigure}{0.31\textwidth}
    \includegraphics[width=\linewidth,clip]{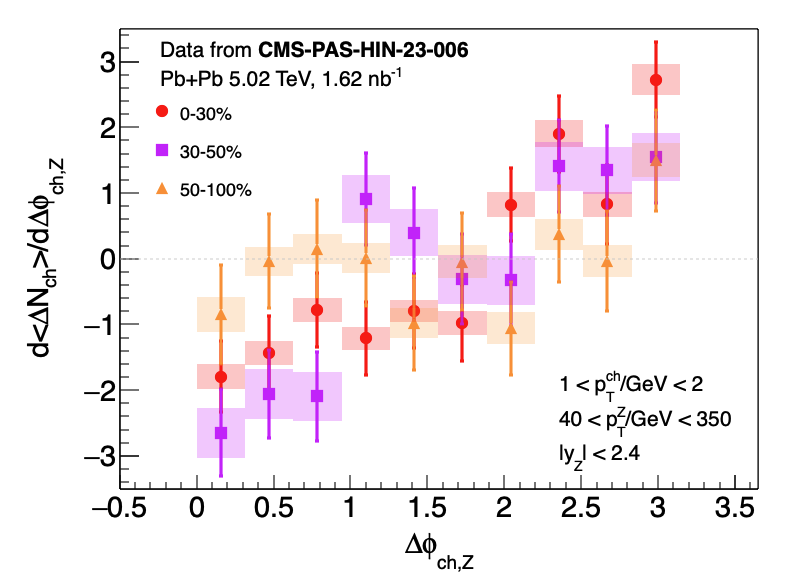}
        \caption{}
    \label{fig:fig5c}
\end{subfigure}
\vspace{-0.5cm}
\caption{\footnotesize (a) ATLAS \cite{ATLAS:2024prm} signal/background ratio in \gajet s as a function of |$\Delta \eta (\mathrm{jet, track})$| for most 0-10\% \PbPb\ collisions and 0.3 < \xJg\ < 0.6 selection. (b) CMS distribution of $\Delta \phi_\mathrm{ch,Z}$ in Z-tagged events ~\cite{CMS:2024fli}. (c) CMS Z-hadron data~\cite{CMS:2024fli} in different centrality selections, overlaid without mirroring.}
\end{figure}

Boson-tagged jets have been proposed as a clean channel to observe the diffusion wake~\cite{Yang:2022nei}. ATLAS and CMS have searched for the diffusion wake in \gajet~\cite{ATLAS:2024prm} and Z-tagged~\cite{CMS:2024fli} events, respectively. ATLAS measures the signal-to-background ratio of jet-track correlations in different ranges of the \xJg\ variable, defined as $p_\mathrm{T}^\mathrm{jet}/p_\mathrm{T}^\gamma$, as a proxy to the energy deposited by the jet in the medium. A stronger diffusion wake signal is expected at $\Delta \eta \sim 0$ and $\Delta \phi \sim \pi$ in events with higher jet suppression. As shown in Fig.~\ref{fig:fig5a}, ATLAS observes a hint of suppression around $|\Delta \eta| = 0$ in the lowest \xJg\ bin. A Monte Carlo sampling analysis yields a best-fit diffusion wake amplitude of 0.5–0.8\% for a wake width in the range 0.5–1.0. This observation is consistent with CoLBT predictions within the 68\% confidence level and sets strong upper limits on the magnitude of the diffusion wake.

CMS presented preliminary results for the search for diffusion wakes in Z+hadron events~\cite{CMS:2024fli} at the conference. Unlike ATLAS, CMS does not require the presence of a back-to-back jet from the boson but instead measures Z-hadron correlations. This approach bypasses limitations imposed by jet definitions and includes contributions from highly suppressed jets, maximizing sensitivity to medium effects. The medium suppression is characterized via centrality selection, providing a looser control compared to \xJg. The spectrum is analyzed for different hadron \pT\ selections as a function of the azimuthal separation between the Z boson and the hadron, $\Delta \phi_\mathrm{ch,Z}$, and their rapidity separation, $\Delta y_\mathrm{ch,Z}$. Fig.~\ref{fig:fig5b} shows the results for the most central (0--30\%) Pb+Pb collisions, mirrored in $\Delta \phi_\mathrm{ch,Z}$. The observable $\mathrm{d}\langle \Delta N_\mathrm{ch}\rangle / \mathrm{d}\Delta \phi_\mathrm{ch,Z}$ is the Z-hadron correlation function, constructed such that its integral over the full phase space in $\Delta \eta$ and $\Delta \phi$ is zero, thereby capturing modifications from both the diffusion wake and the wake aligned with the jet direction. Although the absolute number of particles depleted solely due to the diffusion wake, excluding contributions from the jet-direction wake, is not directly accessible from this observable, theoretical models that include the medium response, including diffusion wake effects, provide the best description of the data in the low-\pT\ region. Interestingly, and perhaps counterintuitively, no clear centrality ordering is observed, with the highest depletion appearing in mid-central collisions, as shown in Fig.~\ref{fig:fig5c}. 

The search for a medium response in boson-tagged events will greatly benefit from the higher statistics collected in Run 3. Both ATLAS and CMS are expected to improve by a factor of 3 to 4 compared to Run 2. 

\vspace{-0.3cm}

\vspace{-0.1cm}
\section{Small systems}

\subsection{Is there evidence of energy loss onset in existing experimental data from small systems?} 

The absence of conclusive evidence for energy loss effects on high-momentum hadrons in small collision systems, such as $p$+A and high-multiplicity \pp\ collisions, remains one of the outstanding challenges to achieving a global understanding of QCD matter formation across different temperature and system-size regimes.
Hard-jet correlations have been exploited both at RHIC and LHC to search for signatures of energy loss. 

\begin{figure}[h]
\centering
\begin{subfigure}{0.31\textwidth}
    \includegraphics[width=\linewidth,clip]{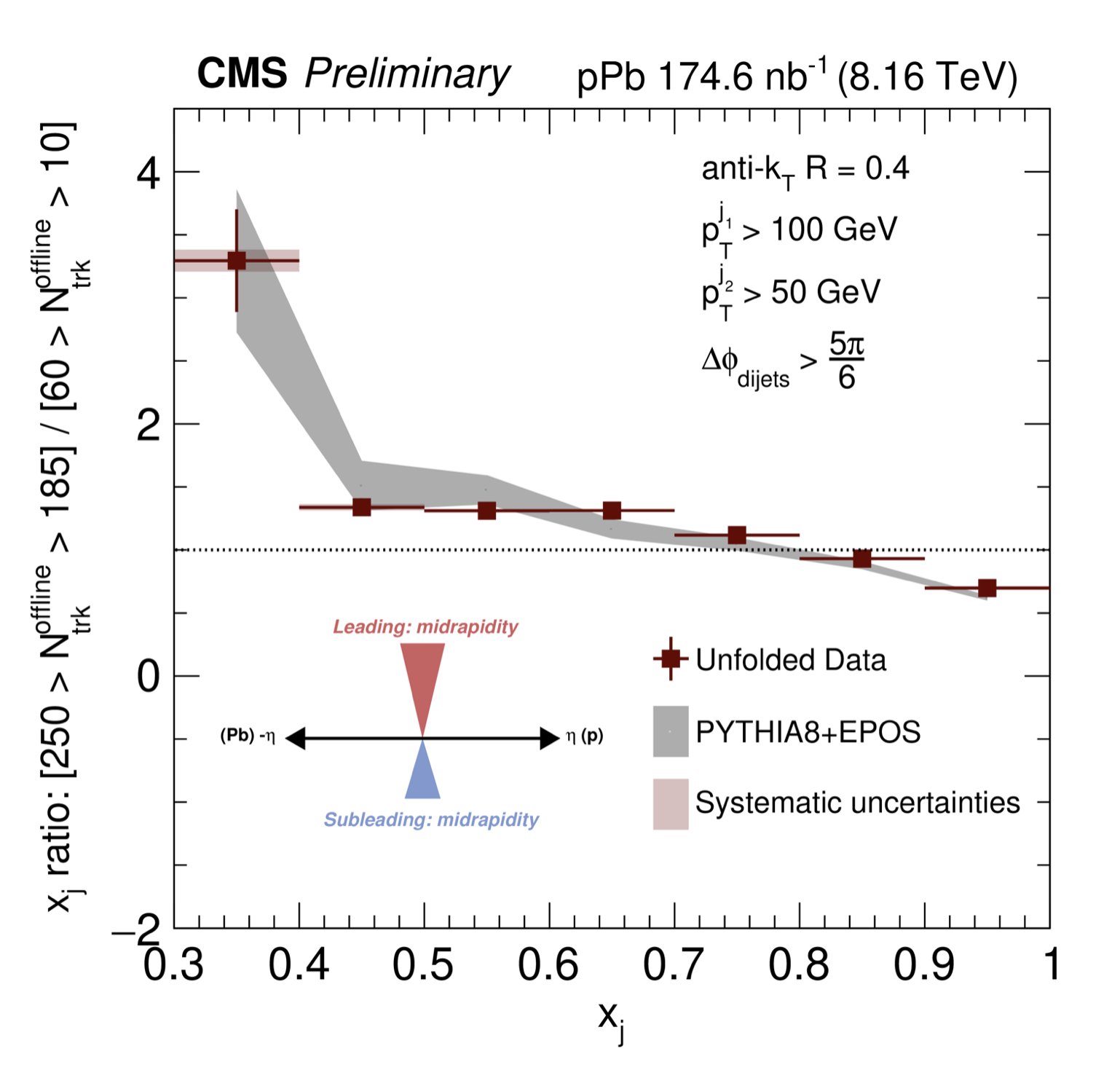}
        \caption{}
    \label{fig:fig6a}
\end{subfigure}
\hfill
\begin{subfigure}{0.4\textwidth}
    \includegraphics[width=\linewidth,clip]{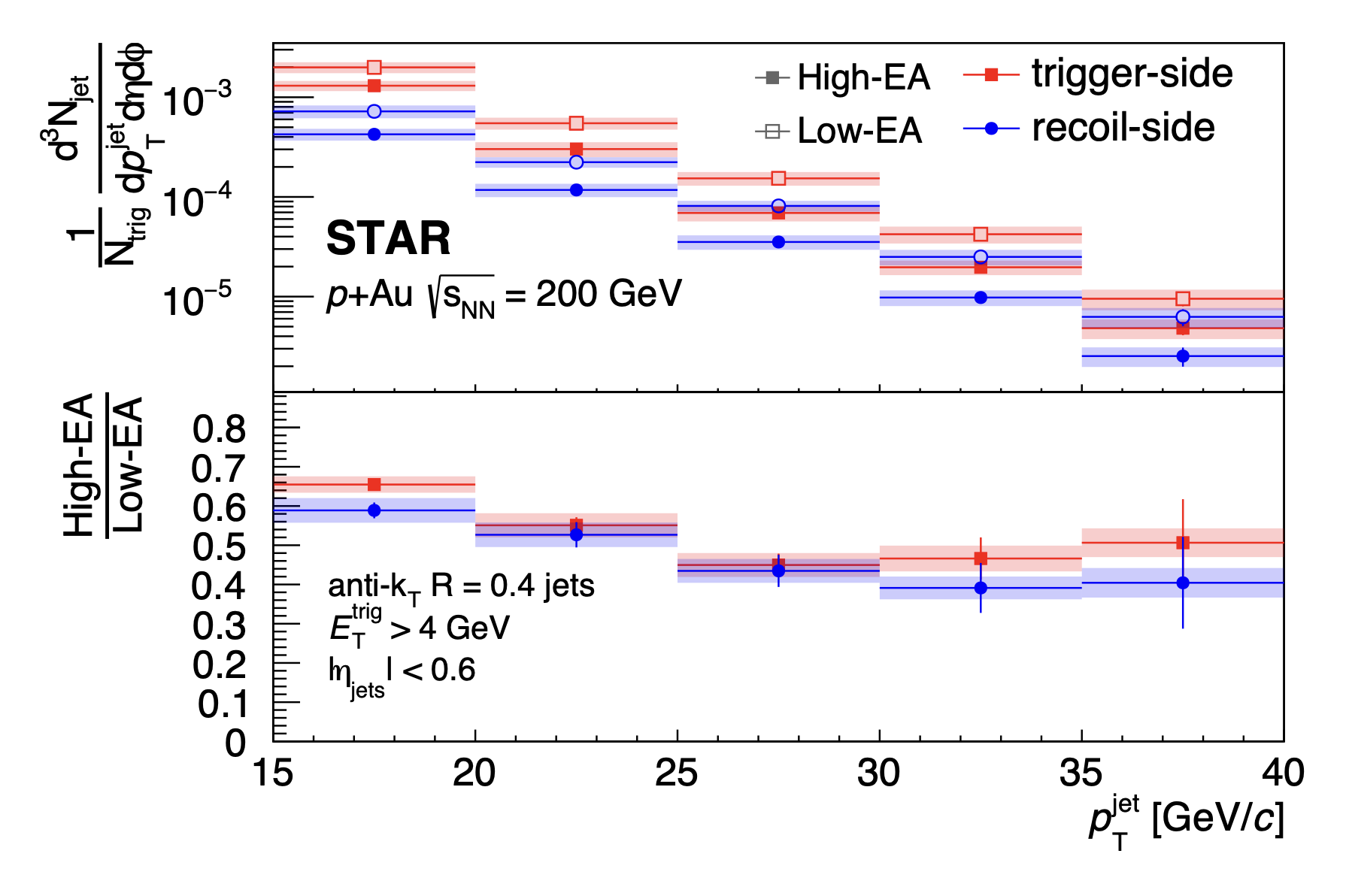}
        \caption{}
    \label{fig:fig6b}
\end{subfigure}
\hfill
\begin{subfigure}{0.2\textwidth}
    \includegraphics[width=\linewidth,clip]{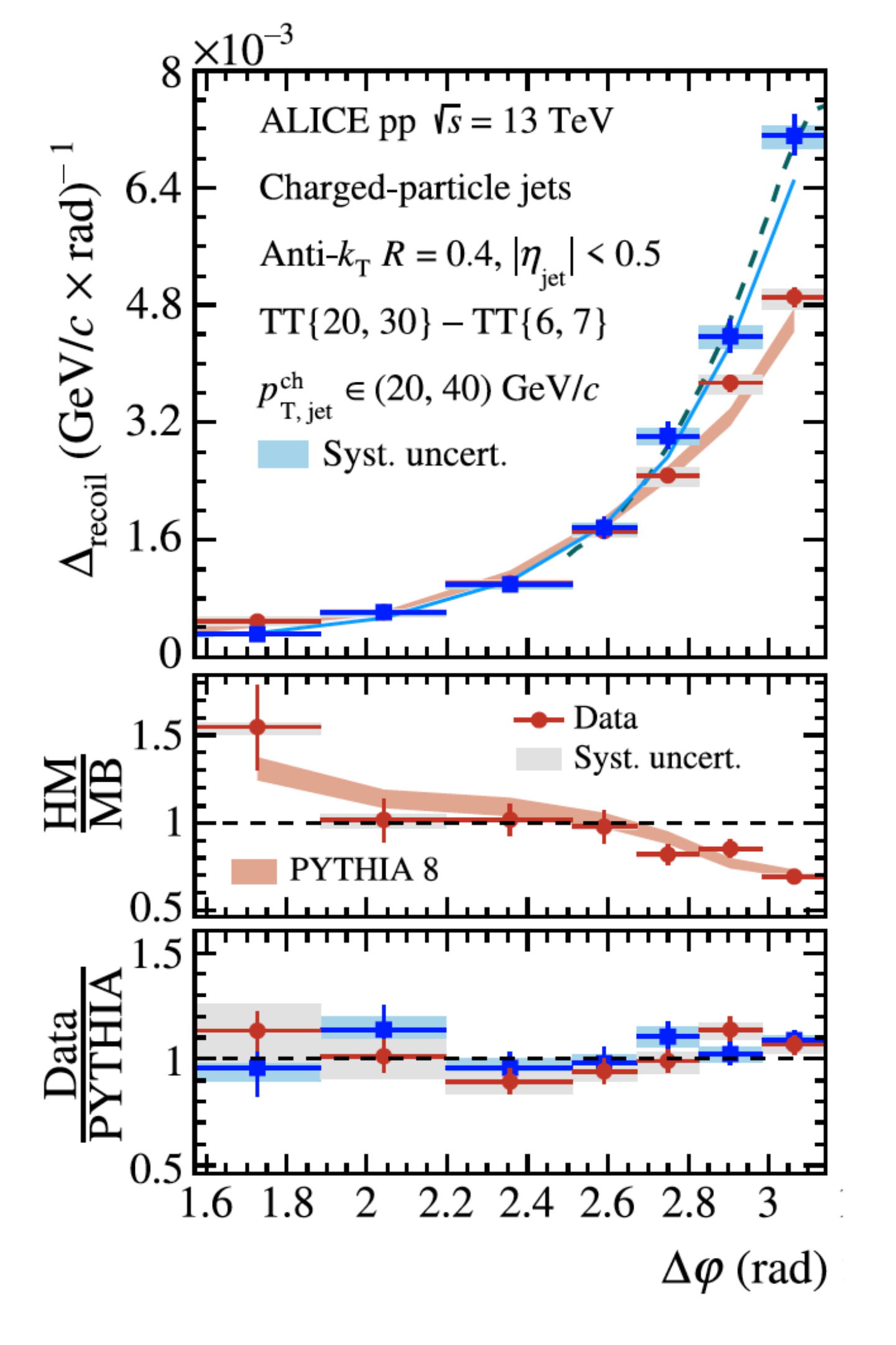}
        \caption{}
    \label{fig:fig6c}
\end{subfigure}
\vspace{-0.5cm}
\caption{\footnotesize (a) CMS \xJ\ ratio between two given track multiplicity selections, compared with \pythia+EPOS MC~\cite{CMS:2024hlf} (b) STAR per-trigger jet spectra for trigger and recoil sides, for both high and low event activity selections~\cite{STAR:2024nwm}. (c) ALICE comparison of hadron+jet azimuthal correlations in high-multiplicity \pp\ events with MB and \pythia~\cite{ALICE:2023plt}.}
\end{figure}

CMS has recently released results on dijet momentum imbalance, \xJ, in \pPb\ collisions, studied as a function of charged track multiplicity and different center-of-mass pseudorapidity configurations \cite{CMS:2024hlf}. The ratio between different track multiplicities is studied as a function of \xJ\ and compared to \pythia+EPOS Monte Carlo (MC) simulations (see Fig.~\ref{fig:fig6a}), which do not include effects related to QGP production. The good agreement between MC and data indicates no evidence of jet quenching signature in small colliding systems in any track multiplicity classes, including the highest one, where strong collectivity effects are observed. 

Semi-inclusive hadron+jet measurements have also been employed to search for energy loss effects at the LHC and RHIC. ATLAS placed strong constraints on parton energy loss in \pPb\ collisions using this approach~\cite{ATLAS:2022iyq}, while at RHIC, the STAR Collaboration recently performed a hadron-jet correlation measurement as a function of event activity classes in $d$+Au collisions. The results, shown in Fig.~\ref{fig:fig6b}, exhibit comparable suppression between high- and low-activity collisions for both the hadron-trigger and jet-recoil sides, with no evidence of path-length dependence. Importantly, this type of observation is not affected by the known issues associated with event activity as a centrality estimator in $p/d$+A collisions~\cite{ATLAS:2023zfx}.

ALICE also used semi-inclusive hadron+jet measurements to search for QGP-like signatures in high-multiplicity \pp\ collisions \cite{ALICE:2023plt}. The results, reported in Fig.~\ref{fig:fig6c}, show a suppression in the back-to-back hadron-jet pairs in high-multiplicity events compared to minimum bias (MB), but not compared to \pythia. This observation was traced back to a selection bias in the data acting toward higher processes, without any quenching implication.  

\vspace{-0.3cm}

\vspace{-0.3cm}

\subsection{How do color fluctuations affect the interpretation of hard scatterings in $p$+A collisions?} 
\begin{figure}[h]
\centering
\begin{subfigure}{0.34\textwidth}
    \includegraphics[width=\linewidth,clip]{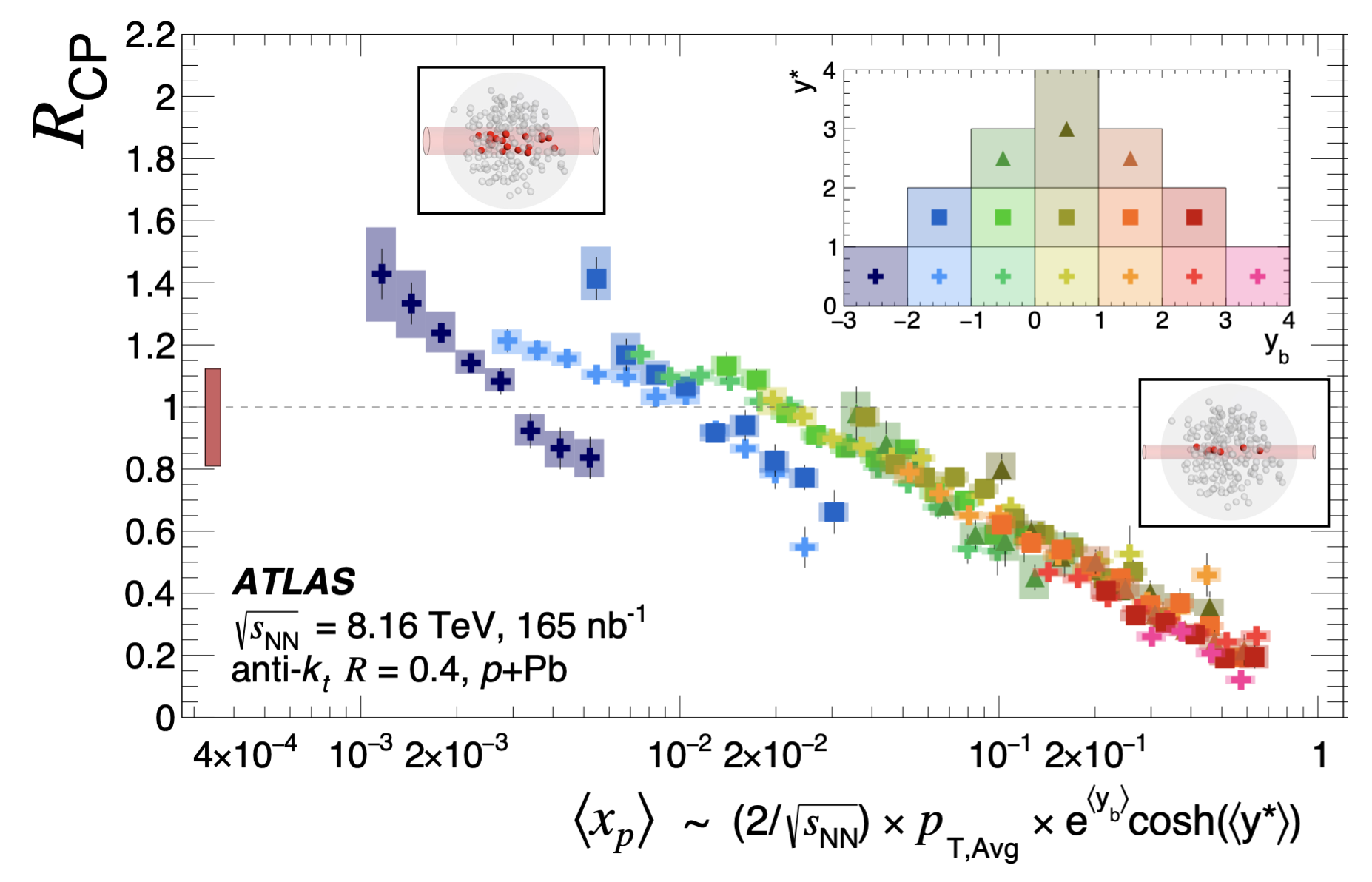}
        \caption{}
    \label{fig:fig7a}
\end{subfigure}
\hfill
\begin{subfigure}{0.3\textwidth}
    \includegraphics[width=\linewidth,clip]{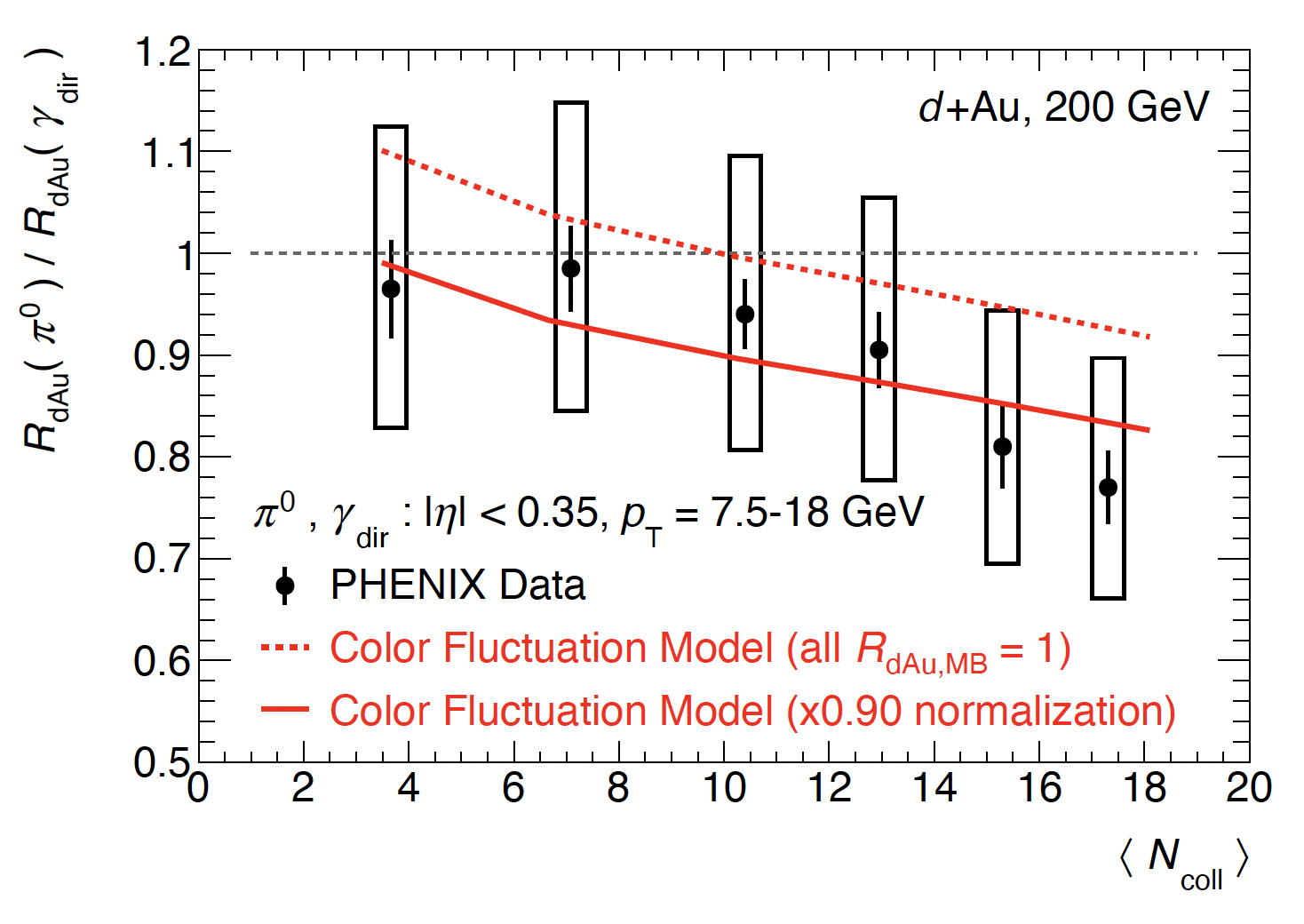}
        \caption{}
    \label{fig:fig7b}
\end{subfigure}
\hfill
\begin{subfigure}{0.3\textwidth}
    \includegraphics[width=\linewidth,clip]{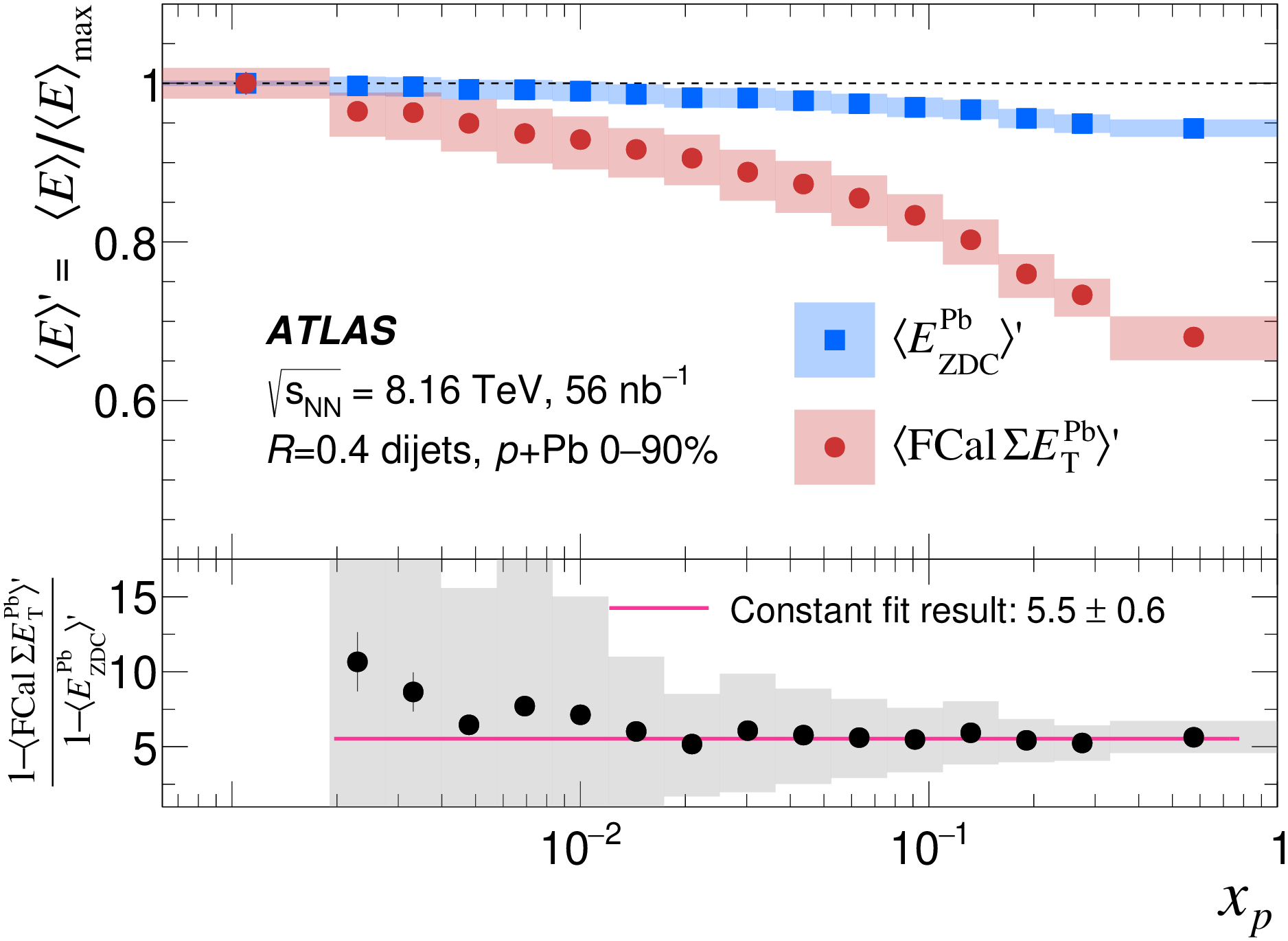}
        \caption{}
    \label{fig:fig7c}
\end{subfigure}
\vspace{-0.5cm}
\caption{\footnotesize (a) ATLAS \RCP\ as a function of the Bjorken-$x$ of the proton, $x_p$~\cite{ATLAS:2023zfx} (b) PHENIX $R_\mathrm{dAu}(\pi^0)/R_\mathrm{dAu}(\gamma_\mathrm{dir})$ from $d$+Au collisions at 200 GeV, compared with results from the color fluctuation model~\cite{Perepelitsa:2024eik}. (c) ATLAS average ZDC (blue circles) and forward calorimeter transverse energy (red squares) as a function of $x_p$~\cite{ATLAS:2025hac}. Both distributions are normalized by their maximum. }
\end{figure}

The relevance of the proton configuration in interpreting event activity in $p$+A collisions was recently highlighted by ATLAS~\cite{ATLAS:2023zfx} using dijets to access the initial-state kinematics. The measurement studied the central-to-peripheral ratio, \RCP, where centrality was defined via the Pb-going forward calorimeter activity, as a function of the Bjorken-$x$ of the colliding partons, $x_p$ and $x_\mathrm{Pb}$. As shown in Fig.~\ref{fig:fig7a}, a strong event activity bias driven by $x_p$ is observed. The suppression of high-activity events at large $x_p$ is consistent with the color fluctuation model of~\cite{Alvioli:2017wou}, which successfully describes inclusive jet results at RHIC and the LHC. In this model, protons containing a large-$x_p$ parton are more compact and interact more weakly, leading to fewer nucleon-nucleon collisions and thus a bias in event activity.

Characterizing color fluctuation effects is essential for the correct interpretation of nuclear collision data. A recent example comes from RHIC, where PHENIX measured the ratio $R_\mathrm{dAu}(\pi^0)/R_\mathrm{dAu}(\gamma_\mathrm{dir})$ in $d$+Au collisions at $\sqrt{s_{\mathrm{NN}}} = 200$~GeV, comparing neutral pion and direct photon yields~\cite{PHENIX:2023dxl}. The analysis aimed to suppress centrality bias by comparing two probes similarly affected by such correctly interpreting, attributing any residual modification to final-state effects. The measured ratio, shown in Fig.~\ref{fig:fig7c}, decreases with increasing number of binary collisions. While this could suggest final-state energy loss, such an interpretation would be in tension with several other RHIC and LHC results. As noted in~\cite{Perepelitsa:2024eik}, $\pi^0$ and $\gamma_\mathrm{dir}$ production are characterized by different deuteron Bjorken-$x$ distributions. The color fluctuation model described in~\cite{Alvioli:2017wou}, run without further tuning, successfully reproduces the observed suppression (red curve in Fig.~\ref{fig:fig7c}), supporting the interpretation that the effect arises from different average projectile configurations and interaction strengths for the $\pi^0$ and $\gamma_\mathrm{dir}$ samples. This reconciles the PHENIX results with other searches for energy loss effects at RHIC and the LHC.

Color fluctuations have also been proposed to influence nuclear break-up in resolved ultra-peripheral~\cite{Alvioli:2024cmd} and \pPb\ collisions~\cite{Alvioli:2025ggv}. At the conference, ATLAS presented the first measurement of nuclear break-up in \pPb\ as a function of hard-scattering kinematics~\cite{ATLAS:2025hac}, using dijets reconstructed across the full calorimeter acceptance. These were correlated with energy deposition in the Pb-going Zero Degree Calorimeter (ZDC) and Forward Calorimeter, as shown in Fig.~\ref{fig:fig7c}. The results provide extensive input for understanding the event activity bias in \pPb\ as a function of proton configuration. While these configurations also affect the ZDC signal, the ZDC energy is approximately six times more resilient to color fluctuation effects than the forward transverse energy.

\vspace{-0.3cm}
\vspace{-0.4cm}

\subsection{How are parton distribution functions modified in the nuclear environment?} 
\begin{figure}[h]
\centering
\begin{subfigure}{0.29\textwidth}
    \includegraphics[width=\linewidth,clip]{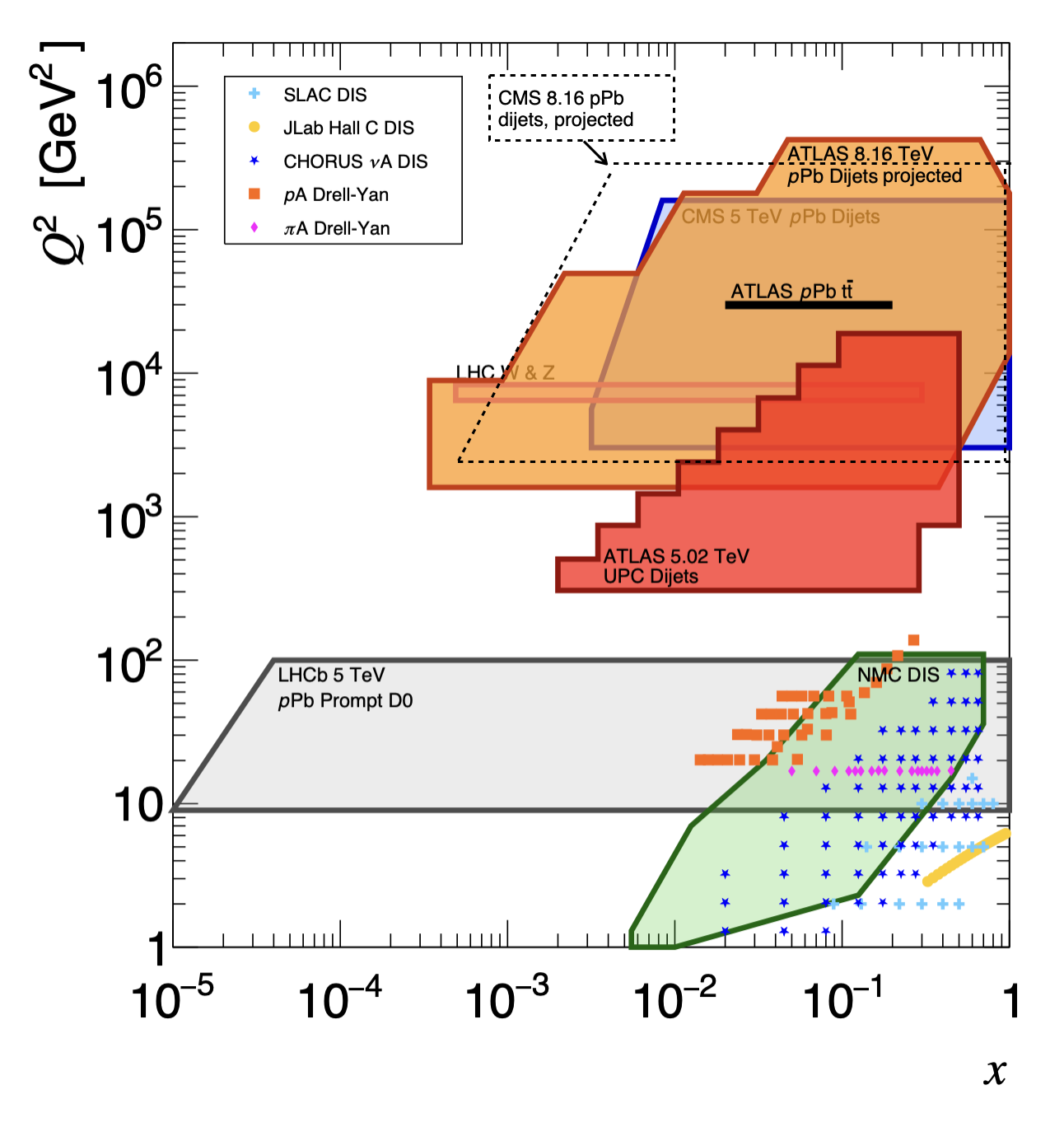}
        \caption{}
    \label{fig:fig8a}
\end{subfigure}
\hfill
\begin{subfigure}{0.25\textwidth}
    \includegraphics[width=\linewidth,clip]{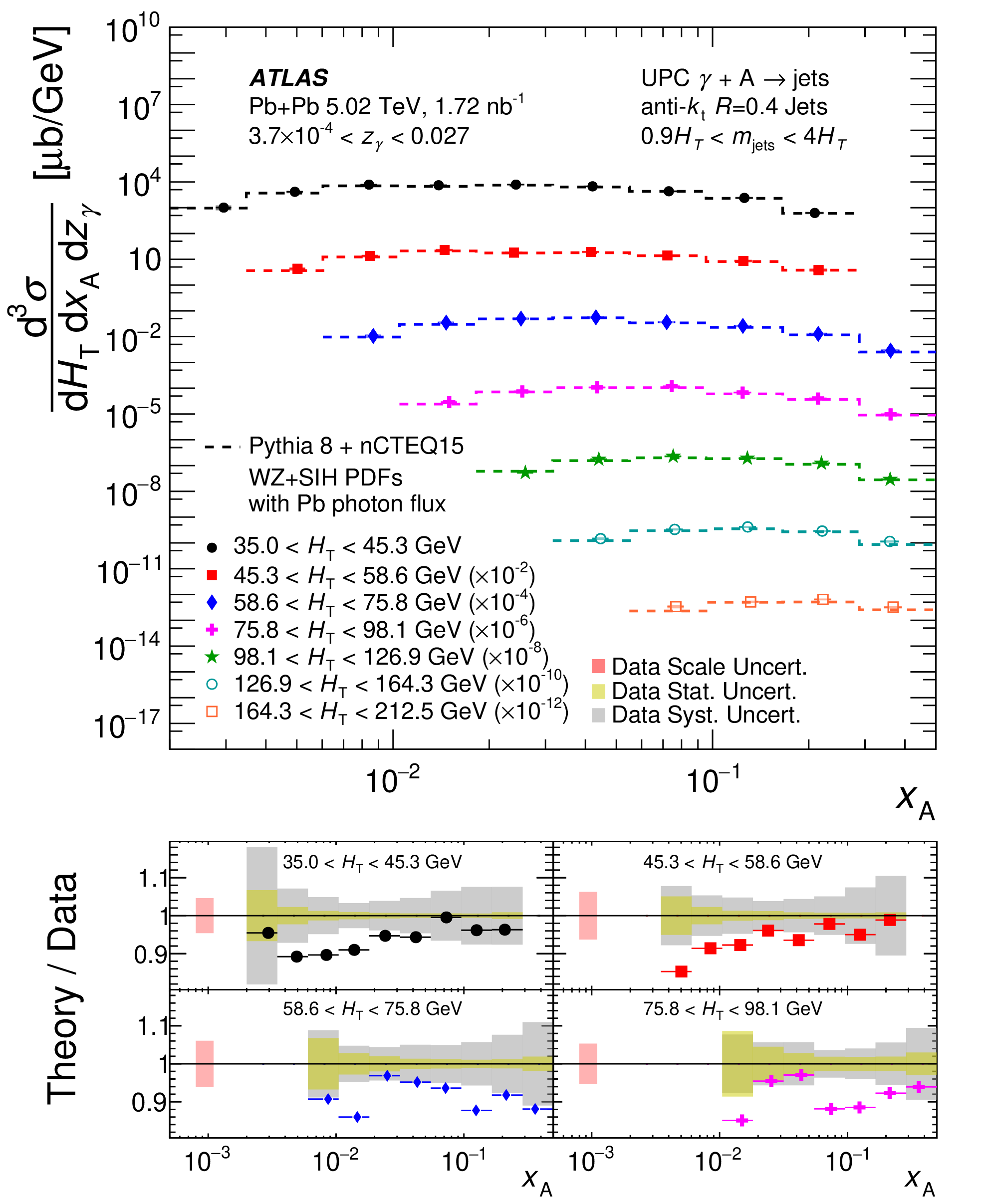}
        \caption{}
    \label{fig:fig8b}
\end{subfigure}
\hfill
\begin{subfigure}{0.38\textwidth}
    \includegraphics[width=\linewidth,clip]{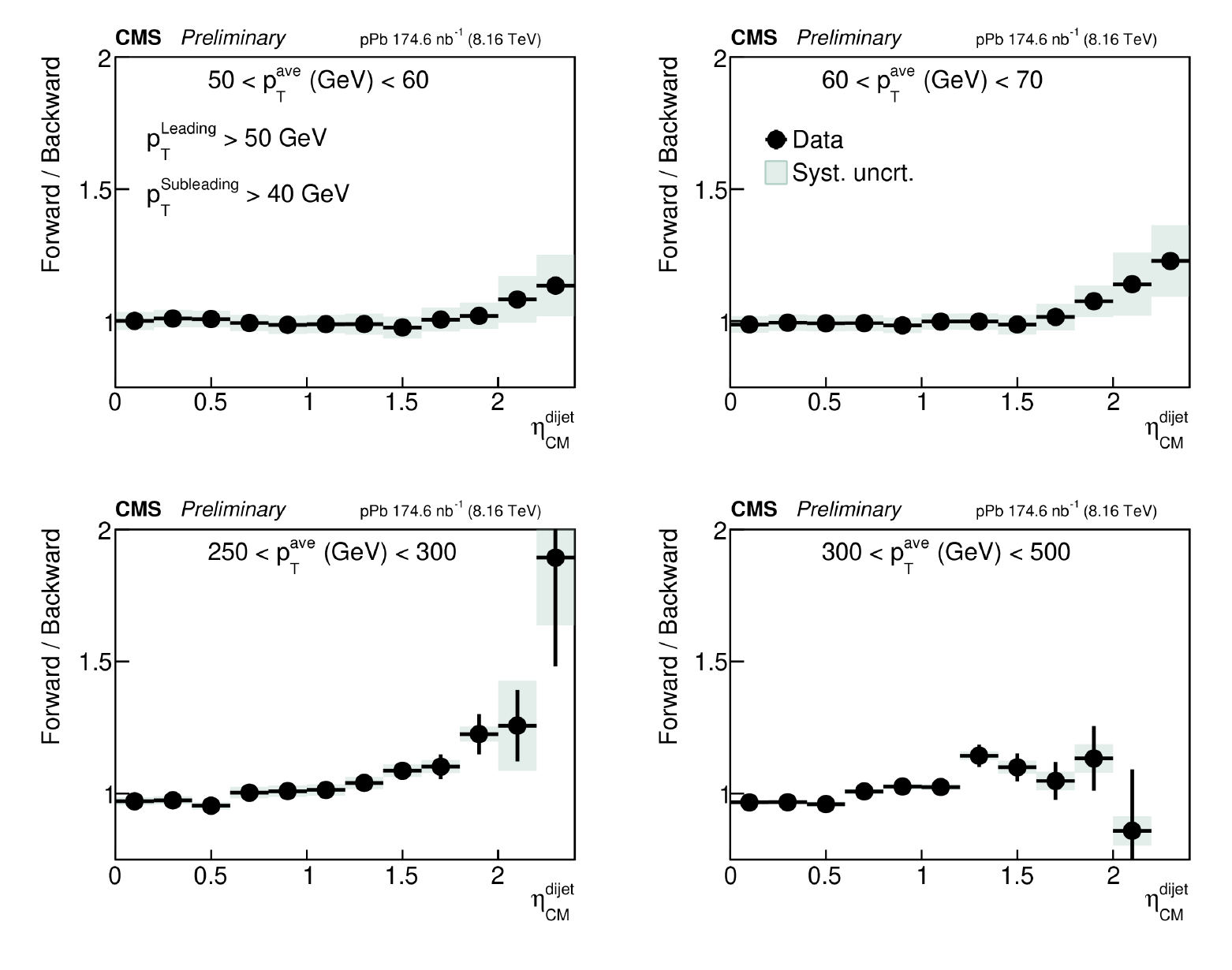}
        \caption{}
    \label{fig:fig8c}
\end{subfigure}
\vspace{-0.5cm}
\caption{\footnotesize (a) ($x$,$Q^2$) coverage of data included in EPPS21 \cite{Eskola:2021nhw} together with new coverage of new results from ATLAS and CMS discussed in this proceeding. (b) ATLAS triple-differential extraction of the dijet cross-section in photo-nuclear events \cite{ATLAS:2024mvt}. (c) CMS forward-to-backward ratio of self-normalized dijet pseudorapidity distributions for different selections of $p_\mathrm{T}^\mathrm{ave}$\cite{CMS:2024idr}.}
\end{figure}

Dijets are also powerful probes for constraining nuclear Parton Distribution Functions (nPDFs). Several recently released results expand nPDF sensitivity across a broad ($x$,$Q^2$) phase space. ATLAS has published the final measurement of the dijet cross-section in photonuclear collisions~\cite{ATLAS:2024mvt}. The asymmetric $\gamma$+A system grants access to an intermediate $Q^2$ region not reachable in symmetric hadronic collisions (Fig.~\ref{fig:fig8a}). The measurement is unfolded in all three analysis dimensions: $H_\mathrm{T}$ (hard scale), $z_\gamma$ (photon resolution proxy), and $x_A$ (momentum fraction of the struck parton). Results are compared to leading-order \pythia\ predictions (Fig.~\ref{fig:fig8b}) as well as to \pythia\ combined with various global nPDF fits.

\pPb\ data have also been used to extract the dijet cross-section and constrain nuclear PDFs (nPDFs). Despite the limitations posed by the absence of a dedicated \pp\ reference at the same center-of-mass energy in Run 2, the high-statistics 8.16TeV \pPb\ dataset collected in 2016 has been used extensively by both ATLAS and CMS to study dijet production. ATLAS published the triple-differential per-event dijet yield~\cite{ATLAS:2023zfx} (orange area in Fig.~\ref{fig:fig8a}), while CMS recently presented preliminary results on self-normalized dijet pseudorapidity distributions across different dijet transverse momentum ($p_\mathrm{T}^\mathrm{ave}$) bins~\cite{CMS:2024idr}. In the absence of a \pp\ baseline, CMS uses forward-to-backward ratios (Fig.~\ref{fig:fig8c}) to reduce sensitivity to free-proton and scale uncertainties, as well as experimental systematics. However, these ratios mix rapidity regions and thus probe correlations between high- and low-$x$ nuclear modifications, rather than isolating the parton momentum fraction. These new results will further constrain nPDFs in a ($x$,$Q^2$) region similar to that accessed by previous CMS measurements using 5.02~TeV data~\cite{CMS:2018jpl} (dashed area in Fig.~\ref{fig:fig8a}).

\vspace{-0.3cm}

\vspace{-0.4cm}

\section{What is next?} 
The LHC Run 3, recently extended to mid-2026, will deliver a substantial increase in HI statistics. ATLAS and CMS are expected to record approximately three to four times more data than in Run 2, boosting statistically demanding measurements, such as those using boson-tagged jets. Thanks to the new continuous stream readout, ALICE is projected to collect about 50 times more data than in Runs 1 and 2 combined, significantly enhancing the precision and scope of its physics program. The LHC will also deliver the first oxygen-oxygen and proton-oxygen collisions in 2025, offering unprecedented opportunities to search for the onset of energy loss through hard-jet correlation measurements and to place the first constraints on oxygen nPDFs. The oxygen program will heavily benefit from a \pp\ reference data collected at the same center-of-mass energy. 

The future is also bright for RHIC as it approaches its transition to the Electron-Ion Collider. The newly built sPHENIX experiment~\cite{Aidala:2012nz} was successfully commissioned in 2024 and is poised to deliver high-precision jet measurements, probing a different quark-gluon mixture compared to the LHC in a complementary \pT\ phase space, partially overlapping with CERN experiments. STAR has also completed its forward upgrade and is currently taking data. Given the remarkable capabilities of these two detectors and the upcoming RHIC shutdown, the community should prioritize securing sufficient Au+Au collisions and a $p$+Au run to fully exploit the new sPHENIX and STAR forward programs before the accelerator ceases operations. 

\section{Acknowledgements} 
The author would like to thank A. Sickles, M. Rybar, B. Gilbert, S. Mohapatra, D. Perepelitsa, Y. Mao, Y. Go, Y. Lee, P. Steinberg, P. Jacobs, C. McGinn, and D. Hangal for useful discussions and input to prepare the talk.


\begin{thebibliography}{}
\bibitem{Yaxian}
Y.Mao, `Jets: experimental overview', Hard Probes 2024, Sept. 26th 2024, \url{https://indi.to/VKmPF}.


%\cite{Caron-Huot:2008zna}
\bibitem{Caron-Huot:2008zna}
S.~Caron-Huot,
%``O(g) plasma effects in jet quenching,''
Phys. Rev. D \textbf{79} (2009), 065039
doi:10.1103/PhysRevD.79.065039
[arXiv:0811.1603 [hep-ph]].
%180 citations counted in INSPIRE as of 26 Apr 2025

%\cite{ATLAS:2023iad}
\bibitem{ATLAS:2023iad}
G.~Aad \textit{et al.} [ATLAS],
%``Comparison of inclusive and photon-tagged jet suppression in 5.02 TeV Pb+Pb collisions with ATLAS,''
Phys. Lett. B \textbf{846} (2023), 138154
doi:10.1016/j.physletb.2023.138154
[arXiv:2303.10090 [nucl-ex]].
%28 citations counted in INSPIRE as of 26 Apr 2025

%\cite{Brewer:2018dfs}
\bibitem{Brewer:2018dfs}
J.~Brewer, J.~G.~Milhano and J.~Thaler,
%``Sorting out quenched jets,''
Phys. Rev. Lett. \textbf{122} (2019) no.22, 222301
doi:10.1103/PhysRevLett.122.222301
[arXiv:1812.05111 [hep-ph]].
%42 citations counted in INSPIRE as of 26 Apr 2025

%\cite{Casalderrey-Solana:2012evi}
\bibitem{Casalderrey-Solana:2012evi}
J.~Casalderrey-Solana, Y.~Mehtar-Tani, C.~A.~Salgado and K.~Tywoniuk,
%``New picture of jet quenching dictated by color coherence,''
Phys. Lett. B \textbf{725} (2013), 357-360
doi:10.1016/j.physletb.2013.07.046
[arXiv:1210.7765 [hep-ph]].

%\cite{ATLAS:2023hso}
\bibitem{ATLAS:2023hso}
G.~Aad \textit{et al.} [ATLAS],
%``Measurement of Suppression of Large-Radius Jets and Its Dependence on Substructure in Pb+Pb Collisions at sNN=5.02\,\,TeV with the ATLAS Detector,''
Phys. Rev. Lett. \textbf{131} (2023) no.17, 172301
doi:10.1103/PhysRevLett.131.172301
[arXiv:2301.05606 [nucl-ex]].
%29 citations counted in INSPIRE as of 25 Apr 2025

%\cite{ATLAS:2022vii}
\bibitem{ATLAS:2022vii}
G.~Aad \textit{et al.} [ATLAS],
%``Measurement of substructure-dependent jet suppression in Pb+Pb collisions at 5.02 TeV with the ATLAS detector,''
Phys. Rev. C \textbf{107} (2023) no.5, 054909
doi:10.1103/PhysRevC.107.054909
[arXiv:2211.11470 [nucl-ex]].
%47 citations counted in INSPIRE as of 26 Apr 2025

%\cite{CMS:2024zjn}
\bibitem{CMS:2024zjn}
A.~Hayrapetyan \textit{et al.} [CMS],
%``Girth and groomed radius of jets recoiling against isolated photons in lead-lead and proton-proton collisions at sNN=5.02~TeV,''
Phys. Lett. B \textbf{861} (2025), 139088
doi:10.1016/j.physletb.2024.139088
[arXiv:2405.02737 [nucl-ex]].
%16 citations counted in INSPIRE as of 26 Apr 2025

\bibitem{MPark}
 [CMS],
CMS-PAS-HIN-21-019.


%\cite{ATLAS:2022zbu}
\bibitem{ATLAS:2022zbu}
G.~Aad \textit{et al.} [ATLAS],
%``Measurements of the suppression and correlations of dijets in Pb+Pb collisions at sNN=5.02 TeV,''
Phys. Rev. C \textbf{107} (2023) no.5, 054908
[erratum: Phys. Rev. C \textbf{109} (2024) no.2, 029901]
doi:10.1103/PhysRevC.107.054908
[arXiv:2205.00682 [nucl-ex]].
%32 citations counted in INSPIRE as of 28 Apr 2025

%\cite{ATLAS:2024jtu}
\bibitem{ATLAS:2024jtu}
G.~Aad \textit{et al.} [ATLAS],
%``Jet radius dependence of dijet momentum balance and suppression in Pb+Pb collisions at 5.02 TeV with the ATLAS detector,''
Phys. Rev. C \textbf{110} (2024) no.5, 054912
doi:10.1103/PhysRevC.110.054912
[arXiv:2407.18796 [nucl-ex]].
%5 citations counted in INSPIRE as of 28 Apr 2025

%\cite{ALICE:2023waz}
\bibitem{ALICE:2023waz}
S.~Acharya \textit{et al.} [ALICE],
%``Measurement of the radius dependence of charged-particle jet suppression in Pb\textendash{}Pb collisions at sNN=5.02TeV,''
Phys. Lett. B \textbf{849} (2024), 138412
doi:10.1016/j.physletb.2023.138412
[arXiv:2303.00592 [nucl-ex]].
%42 citations counted in INSPIRE as of 28 Apr 2025

%\cite{ALICE:2023qve}
\bibitem{ALICE:2023qve}
S.~Acharya \textit{et al.} [ALICE],
%``Observation of Medium-Induced Yield Enhancement and Acoplanarity Broadening of Low-pT Jets from Measurements in pp and Central Pb-Pb Collisions at sNN=5.02\,\,TeV,''
Phys. Rev. Lett. \textbf{133} (2024) no.2, 022301
doi:10.1103/PhysRevLett.133.022301
[arXiv:2308.16131 [nucl-ex]].
%17 citations counted in INSPIRE as of 28 Apr 2025

%\cite{He:2024rcv}
\bibitem{He:2024rcv}
Y.~He, M.~Nie, S.~Cao, R.~Ma, L.~Yi and H.~Caines,
%``Deciphering yield modification of hadron-triggered semi-inclusive recoil jets in heavy-ion collisions,''
Phys. Lett. B \textbf{854} (2024), 138739
doi:10.1016/j.physletb.2024.138739
[arXiv:2401.05238 [nucl-th]].
%5 citations counted in INSPIRE as of 28 Apr 2025

%\cite{Yang:2022nei}
\bibitem{Yang:2022nei}
Z.~Yang, T.~Luo, W.~Chen, L.~G.~Pang and X.~N.~Wang,
%``3D Structure of Jet-Induced Diffusion Wake in an Expanding Quark-Gluon Plasma,''
Phys. Rev. Lett. \textbf{130} (2023) no.5, 052301
doi:10.1103/PhysRevLett.130.052301
[arXiv:2203.03683 [hep-ph]].
%39 citations counted in INSPIRE as of 28 Apr 2025

%\cite{ATLAS:2024prm}
\bibitem{ATLAS:2024prm}
G.~Aad \textit{et al.} [ATLAS],
%``Search for the jet-induced diffusion wake in the quark-gluon plasma via measurements of jet-track correlations in photon-jet events in Pb+Pb collisions at $\sqrt{\mathrm{s}_{\mathrm{NN}}}$ = 5.02 TeV with the ATLAS detector,''
Phys. Rev. C \textbf{111} (2025), 044909
doi:10.1103/PhysRevC.111.044909
[arXiv:2408.08599 [nucl-ex]].
%3 citations counted in INSPIRE as of 28 Apr 2025

%\cite{CMS:2024fli}
\bibitem{CMS:2024fli}
 [CMS],
%``Evidence of the medium response with Z-hadron correlations in PbPb and pp collisions at sqrt(sNN) = 5.02 TeV,''
CMS-PAS-HIN-23-006.
%2 citations counted in INSPIRE as of 28 Apr 2025

%\cite{CMS:2024hlf}
\bibitem{CMS:2024hlf}
 [CMS],
%``Search for jet quenching signature using transverse momentum balance in high-multiplicity pPb collisions at the CMS detector,''
CMS-PAS-HIN-23-010.
%0 citations counted in INSPIRE as of 29 Apr 2025

%\cite{ATLAS:2022iyq}
\bibitem{ATLAS:2022iyq}
G.~Aad \textit{et al.} [ATLAS],
%``Strong Constraints on Jet Quenching in Centrality-Dependent p+Pb Collisions at 5.02~TeV from ATLAS,''
Phys. Rev. Lett. \textbf{131} (2023) no.7, 072301
doi:10.1103/PhysRevLett.131.072301
[arXiv:2206.01138 [nucl-ex]].
%31 citations counted in INSPIRE as of 29 Apr 2025

%\cite{STAR:2024nwm}
\bibitem{STAR:2024nwm}
M.~Abdulhamid \textit{et al.} [STAR],
%``Correlations of event activity with hard and soft processes in p+Au collisions at sNN=200 GeV at the RHIC STAR experiment,''
Phys. Rev. C \textbf{110} (2024) no.4, 044908
doi:10.1103/PhysRevC.110.044908
[arXiv:2404.08784 [nucl-ex]].
%4 citations counted in INSPIRE as of 29 Apr 2025

%\cite{ATLAS:2023zfx}
\bibitem{ATLAS:2023zfx}
G.~Aad \textit{et al.} [ATLAS],
%``Measurement of the Centrality Dependence of the Dijet Yield in p+Pb Collisions at sNN=8.16\,\,TeV with the ATLAS Detector,''
Phys. Rev. Lett. \textbf{132} (2024) no.10, 102301
doi:10.1103/PhysRevLett.132.102301
[arXiv:2309.00033 [nucl-ex]].
%11 citations counted in INSPIRE as of 29 Apr 2025

%\cite{ALICE:2023plt}
\bibitem{ALICE:2023plt}
S.~Acharya \textit{et al.} [ALICE],
%``Search for jet quenching effects in high-multiplicity pp collisions at $ \sqrt{s} $ = 13 TeV via di-jet acoplanarity,''
JHEP \textbf{05} (2024), 229
doi:10.1007/JHEP05(2024)229
[arXiv:2309.03788 [hep-ex]].
%18 citations counted in INSPIRE as of 29 Apr 2025

%\cite{Alvioli:2017wou}
\bibitem{Alvioli:2017wou}
M.~Alvioli, L.~Frankfurt, D.~Perepelitsa and M.~Strikman,
%``Global analysis of color fluctuation effects in proton\textendash{} and deuteron\textendash{}nucleus collisions at RHIC and the LHC,''
Phys. Rev. D \textbf{98} (2018) no.7, 071502
doi:10.1103/PhysRevD.98.071502
[arXiv:1709.04993 [hep-ph]].
%32 citations counted in INSPIRE as of 29 Apr 2025

%\cite{PHENIX:2023dxl}
\bibitem{PHENIX:2023dxl}
N.~J.~Abdulameer \textit{et al.} [PHENIX],
%``Disentangling Centrality Bias and Final-State Effects in the Production of High-pT Neutral Pions Using Direct Photon in d+Au Collisions at sNN=200\,\,GeV,''
Phys. Rev. Lett. \textbf{134} (2025) no.2, 022302
doi:10.1103/PhysRevLett.134.022302
[arXiv:2303.12899 [nucl-ex]].
%18 citations counted in INSPIRE as of 30 Apr 2025

%\cite{ATLAS:2025hac}
\bibitem{ATLAS:2025hac}
G.~Aad \textit{et al.} [ATLAS],
%``Characterization of nuclear breakup as a function of hard-scattering kinematics using dijets measured by ATLAS in $p$+Pb collisions,''
[arXiv:2504.02638 [nucl-ex]].
%1 citations counted in INSPIRE as of 29 Apr 2025


%\cite{Perepelitsa:2024eik}
\bibitem{Perepelitsa:2024eik}
D.~V.~Perepelitsa,
%``Contribution to differential \ensuremath{\pi}0 and \ensuremath{\gamma}dir modification in small systems from color fluctuation effects,''
Phys. Rev. C \textbf{110} (2024) no.1, L011901
doi:10.1103/PhysRevC.110.L011901
[arXiv:2404.17660 [nucl-th]].
%6 citations counted in INSPIRE as of 30 Apr 2025

%\cite{Alvioli:2024cmd}
\bibitem{Alvioli:2024cmd}
M.~Alvioli, V.~Guzey and M.~Strikman,
%``Slicing Pomerons in ultraperipheral collisions using forward neutrons from nuclear breakup,''
Phys. Rev. C \textbf{110} (2024) no.2, 025205
doi:10.1103/PhysRevC.110.025205
[arXiv:2402.19060 [hep-ph]].
%6 citations counted in INSPIRE as of 30 Apr 2025

%\cite{Alvioli:2025ggv}
\bibitem{Alvioli:2025ggv}
M.~Alvioli, V.~Guzey and M.~Strikman,
%``Probing fluctuating protons using forward neutrons in soft and hard inelastic proton-nucleus scattering,''
[arXiv:2504.07514 [hep-ph]].
%0 citations counted in INSPIRE as of 30 Apr 2025

%\cite{ATLAS:2024mvt}
\bibitem{ATLAS:2024mvt}
G.~Aad \textit{et al.} [ATLAS],
%``Measurement of photonuclear jet production in ultraperipheral Pb+Pb collisions at sNN=5.02\,\,TeV with the ATLAS detector,''
Phys. Rev. D \textbf{111} (2025) no.5, 052006
doi:10.1103/PhysRevD.111.052006
[arXiv:2409.11060 [nucl-ex]].
%10 citations counted in INSPIRE as of 30 Apr 2025

%\cite{CMS:2018jpl}
\bibitem{CMS:2018jpl}
A.~M.~Sirunyan \textit{et al.} [CMS],
%``Constraining gluon distributions in nuclei using dijets in proton-proton and proton-lead collisions at $\sqrt{s_{_\mathrm{NN}}} =$ 5.02 TeV,''
Phys. Rev. Lett. \textbf{121} (2018) no.6, 062002
doi:10.1103/PhysRevLett.121.062002
[arXiv:1805.04736 [hep-ex]].
%64 citations counted in INSPIRE as of 30 Apr 2025

%\cite{CMS:2024idr}
\bibitem{CMS:2024idr}
 [CMS],
%``Constraining nPDFs using dijet production in pPb collisions at 8.16 TeV with the CMS experiment,''
CMS-PAS-HIN-24-014.
%0 citations counted in INSPIRE as of 30 Apr 2025

%\cite{Eskola:2021nhw}
\bibitem{Eskola:2021nhw}
K.~J.~Eskola, P.~Paakkinen, H.~Paukkunen and C.~A.~Salgado,
%``EPPS21: a global QCD analysis of nuclear PDFs,''
Eur. Phys. J. C \textbf{82} (2022) no.5, 413
doi:10.1140/epjc/s10052-022-10359-0
[arXiv:2112.12462 [hep-ph]].
%142 citations counted in INSPIRE as of 30 Apr 2025

%\cite{Aidala:2012nz}
\bibitem{Aidala:2012nz}
C.~Aidala, \textit{et al.}
%``sPHENIX: An Upgrade Concept from the PHENIX Collaboration,''
[arXiv:1207.6378 [nucl-ex]].
%67 citations counted in INSPIRE as of 29 Apr 2025

\end{thebibliography}
\end{document}